# Keeping Track of User Steering Actions in Dynamic Workflows

Renan Souza[1,2], Vítor Silva[1*], Jose J. Camata[3],
Alvaro L. G. A. Coutinho[1], Patrick Valduriez[4], Marta Mattoso[1]

[1]COPPE/Federal University of Rio de Janeiro, Brazil
[2]IBM Research
[3]Department of Computer Science, Federal University of Juiz de Fora
[4]Inria and LIRMM, Montpellier, France

**Abstract**

In long-lasting scientific workflow executions in HPC machines, computational scientists (the *users* in this work) often need to fine-tune several workflow parameters. These tunings are done through user steering actions that may significantly improve performance (*e.g.*, reduce execution time) or improve the overall results. However, in executions that last for weeks, users can lose track of what has been adapted if the tunings are not properly registered. In this work, we build on provenance data management to address the problem of tracking online parameter fine-tuning in dynamic workflows steered by users. We propose a lightweight solution to capture and manage provenance of the steering actions online with negligible overhead. The resulting provenance database relates tuning data with data for domain, dataflow provenance, execution, and performance, and is available for analysis at runtime. We show how users may get a detailed view of the execution, providing insights to determine when and how to tune. We discuss the applicability of our solution in different domains and validate its ability to allow for online capture and analyses of parameter fine-tunings in a real workflow in the Oil and Gas industry. In this experiment, the user could determine which tuned parameters influenced simulation accuracy and performance. The observed overhead for keeping track of user steering actions at runtime is less than 1% of total execution time.

## 1. Introduction

In typical High-Performance Computing (HPC) scientific workflows, or workflows for short, computational scientists (the *users* in this work) need to set-up several configuration parameters. These users are specialists in computational models or simulations to solve complex physical problems. They select initial values for the parameters based on their domain expertise. These parameters include solver options, tolerances, and error thresholds. Because of the exploratory nature of those computations, it is hard to determine, before the execution, which configuration values will work best, even for the most experienced users. For this reason, dynamic workflows (*i.e.*, workflows that can be changed at runtime) allow for fine-tunings of specific parameters [1]. These workflow dynamic adaptations are known as user steering actions. After the initial setups, the user starts the computation and, based on online intermediate data analysis, fine-tunes data. Online intermediate data analysis is supported by monitoring tools [1,2] and user steering involves several actions, such as

* Vítor Silva is currently at Dell EMC.

defining steering points, checking-points and rolling-back, refining loop conditions, reducing datasets, modification of filter conditions, and parameter tuning [3]. Parameter tuning is by far the mostly supported one by computational steering solutions [1,2,4–10]. Due to the large number of parameters and combinations of values, uncontrolled parameter fine tunings may lead to rework and difficulties in overall data analysis.

In iterative workflows, where several parameters drive each iteration, analyzing results from initial executions may suggest better settings for the parameters in the following ones. For example, training deep neural networks in large datasets is complex, time consuming, demands parallel computation, and user steering. Often, machine learning experts fine tune the training hyperparameters (*e.g.*, learning rate, batch size, number of training iterations) based on the evolution of the performance of the model and the training time [11]. In Astronomy applications, users may set up data and input parameters to assemble custom mosaics of the sky. During the execution, data analyses may identify that certain input parameters produced images with poor image resolution or quality, making it harder to identify an interesting celestial object. Such parameters can be modified at runtime. In Computational Fluid Dynamics applications, users tune several parameters of the underlying numerical methods [12]. As a result, fine tunings can generate major improvements in performance, resource consumption, and quality of results [13]. Despite the current initiatives to support computational steering in large-scale scientific computing, such as surveyed in [1,2], it remains an open problem [13,14].

Computational steering solutions [1,2,4–10] allow for steering actions. Capturing and registering user steering data (*e.g.*, why the user decided to tune, what were the values before and after the tuning, who and when tuned), relating them to other relevant data (*e.g.*, domain-specific strategic values, execution state of the simulation when the tuning happened, performance data), and allowing all these data to be efficiently integrated and queried at runtime deliver important advantages to the user. They contribute to online data analysis and data-driven decisions. On the other hand, failing to capture steering data has several disadvantages. It may compromise experiment reproducibility and results' reliability as users hardly remember what and how dataflow elements were modified (especially modifications in early stages), and what happened to the execution because of a specific adaptation. This is more critical when users adapt several times in long experiments, which may last for weeks. In addition to losing track of changes, one misses opportunities to learn from the adaptation data (*i.e.*, data generated when humans adapt a certain dataflow element) with the associated dataflow. For example, by registering adaptation data, one may query the data and discover that when parameters are changed to certain range of values, the output result improves by a defined amount. Moreover, opportunities to use the data for AI-based assistants recommending on what to adapt next, based on a database of adaptations, are lost.

In this work, we build on provenance data management to address the problem of keeping track of online parameter fine-tuning in dynamic workflows steered by users. In two recent surveys [13,14], the authors report that solutions for online provenance management and human-in-the-loop of workflows are lacking. To capture and manage provenance of the steering actions online, we consider three steps and their challenges:

**Challenge 1: Online data analysis.** Online data analysis is essential for monitoring, debugging and user steering. In workflows with several parameters to be setup, the user needs to inspect the evolution of results, correlate them with specific input parameter values, and determine which input value is influencing specific outputs [2]. Otherwise, the user will hardly know what or when it should be tuned. According to a recent report [13], current online data analysis solutions are not aware of parameter combinations and their relations with output values.

**Challenge 2: Register the steering action.** Several systems support user steering and parameter fine-tuning [1,2,4–10], but none of them track the steering actions. Not tracking the steering actions jeopardizes the experiment reproducibility. In [13], the authors also state that it is still a challenge to develop a sufficiently descriptive and detailed provenance model to represent steering to enable processing, optimization, validation, interpretation, and reproducibility.

**Challenge 3: Evaluate the steering action.** Enabling online data analyses aware of human adaptations supports data-driven decisions, retrieval of recorded human actions, and understanding of how they relate to the workflow execution status (*e.g.*, how a user action impacts the processing time?). To evaluate the adaptations, the user needs an online query support to access who, when, what was adapted, and how the steering action relates to other data.

In previous works [15,16], we show how applications can benefit from online analysis for steering, supporting Challenge 1, but we are not aware of other works that have addressed the latter two challenges. In [17], we presented an abstract with preliminary ideas to investigate the potential for registering steering actions. In this paper, we formalize steering actions and propose DfAdapter, a lightweight solution to capture and analyze online steering actions in workflows.

To evidence the benefits of keeping track of online parameter fine-tuning, we explore a motivating real case study in the Oil and Gas industry. There are over 50 configuration parameters and their values have a direct impact on the simulation. With the aid of online data analysis, the user can understand which parameters are needed to be tuned and do the adjustment, often several times. For example, the user may identify online regions of interest, which should have more iterations and higher resolution, and regions that can be processed in a lower resolution. This requires adapting several times rather than choosing one single best configuration for the whole workflow execution.

There are several advantages in using DfAdapter with an HPC machine to control online the fine-tuning of the workflow. First, users can evaluate which specific parameter and which ranges of values they modified at runtime led to reduction of memory consumption. Second, DfAdapter helps the user with more data and ways to query these data to allow for better data-driven decisions. More specifically, by using data captured by DfAdapter, the users can verify which parameters were modified, at which iteration in the loop, and when (in time) their steering actions caused the simulation execution time to be reduced by a certain amount, leading their simulation to finish faster, with results they found satisfying. Finally, we observe that the overhead added by DfAdapter for provenance and steering action tracking account for less than 1% of the total execution time.

**Paper organization.** Section 2 presents our motivating case study work. Section 3 presents related work. Section 4 presents our approach for tracking online steering in dataflows. Section 5 presents DfAdapter. In Section 6, we discuss our approach applied to two real-world scientific workflows in the Astronomy and in the Oil and Gas domains. Section 7 shows the experiments. Section 8 concludes.

## 2. Motivating Case Study

The case study explored in this paper is based on a real Computational Fluid Dynamics application in the Oil and Gas domain, called libMesh-sedimentation [18]. It is a simulation solver, implemented in C++ with source code available on GitHub [19], built on top of a widely used parallel fine element framework, libMesh [20], which supports parallel simulation of multiscale, multiphysics applications. libMesh interfaces with several libraries for Computational Science and Engineering applications (*e.g.*, PeTSc, Metis, Parmetis, LAPACK). Also, scientific visualization tools like ParaView [21], are typically used in these applications to gain insight from the computations. In this class of applications, users need to set-up the goals of the computation, and parameters for the numerical methods. Examples of parameters are tolerances for linear and nonlinear solvers, number of levels for mesh adaptation, tolerances for space and time error estimates, etc. These parameters have a direct influence on the accuracy and simulation costs, and bad choices may lead to inaccuracies and even to a simulation crash. As an example, the number of finite elements predicted by the mesh adaptation procedure may exceed the memory available in a processor, and the simulation is halted with an error message. In simulations with complex dynamics, it is often very difficult to set-up a priori a maximum number of finite elements per core that will guarantee the necessary accuracy without exhausting the available resources. Thus, Quantities of Interest (QoIs) like number of finite elements predicted must be tracked and analyzed during execution. The resulting application can be seen as an iterative workflow, as illustrated in Figure 1.

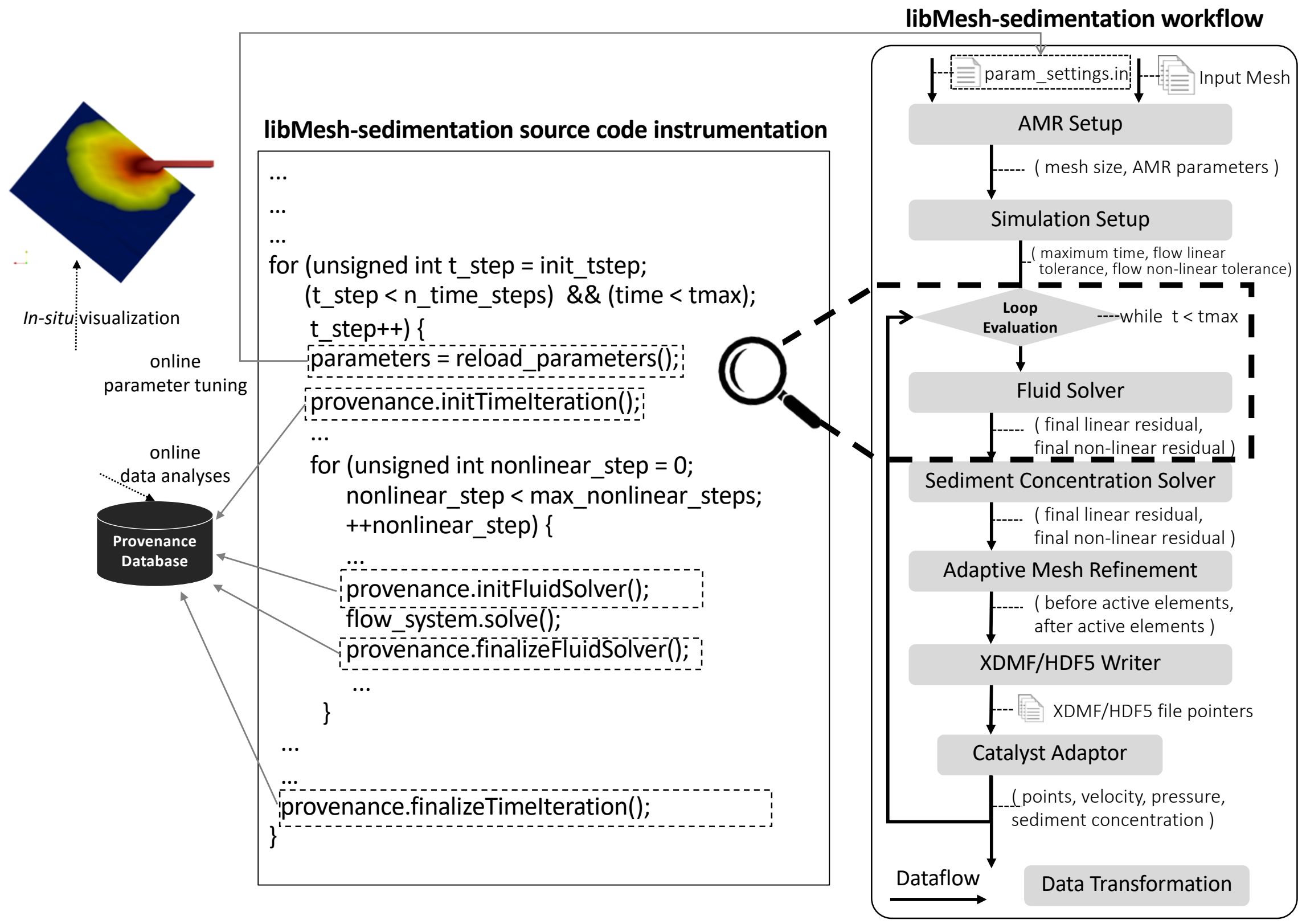

**Figure 1. Keeping track of user steering in the libMesh-sedimentation (adapted from previous work [18]).**

In libMesh-sedimentation, users identify a workflow within the simulation code. They instrument the code to capture monitoring data, which are relevant data for online analysis, and add a steering point (after the `for` loop in `time < tmax`, in Figure 1). Monitoring data captured are stored in a provenance database that follows W3C PROV standards and, in this work, we introduce the track of steering actions to be registered and properly related in a provenance database.

To be able to accomplish this, we present a methodology that describes the steps needed to register and evaluate steering actions. Some of these steps occur offline, before the execution starts, whereas others occur online, during the execution. The offline steps are mainly related to invoking DfAdapter services at the user code, which is a common practice in scientific applications [2]. The user code works as a script, which automates the execution of tasks and often does calls to parallel libraries or other services. We extended our previous methodology [22] to add user steering support. Figure 2 summarizes all high-level steps for enabling workflow steering.

| | | |
|---|---|---|
| 1 | Identify a workflow in the user code | Before execution |
| 2 | Add monitoring points in the code | |
| 3 | Add steering points in the code | |
| 4 | Online data analysis | During execution |
| 5 | Execute a steering action | |
| 6 | Steering action tracking | |
| 7 | Steering action analysis | |

**Figure 2. Methodology for workflow steering.**

**Methodology.** In Step 1, users identify inputs and outputs of relevant parts of their code to form a workflow of chained activities with a dataflow between activities. These inputs and outputs are often domain-specific relevant data (like QoIs) for the users so they can monitor the evolution of the simulation, analyze intermediate data, and understand partial results during the long run. Also, users specify the initial settings and input datasets. In Step 2, users insert service calls in the workflow code to add *monitoring points*. In monitoring points, input and output data elements in the dataflow are specified so they can be captured for monitoring and online data analysis. In Step 3, users identify parts of the code that can be dynamically modified at runtime and add *steering points* in those parts. Steering points should be added in safe points of the code to avoid execution or data inconsistencies. Usually, users know where to add steering points. A typical example occurs in iterative workflows where each new iteration is an opportunity to redefine parameters or input datasets preset beforehand. In this case, a steering point is added in the beginning of the iteration. Each iteration is often executed as a whole. When a user steers, the steering will take effect only at the next iteration, rather than changing values during an iteration. This helps to make data and execution consistent to what the user decided to steer during the iteration.

After these three initial offline steps, the workflow is submitted to parallel execution in an HPC machine. In Step 4, those monitoring data specified at Step 2 are captured and can be analyzed online. In Step 5, based on the analyses, users may decide to execute a steering action. In Step 6, the system tracks steering actions and relates to the data being captured. Finally, in Step 7, users analyze the consequences of their actions relating to domain-specific relevant data and execution data (*e.g.*, time taken to execute a processing). We highlight that, to the best of our knowledge, Steps 6 and 7 are not supported in Computational Steering systems [1,2,4–10].

## 3. Related Work in Computational Steering in Scientific Workflows

In a recent survey [23], the authors discuss past, present, and future of scientific workflows. As a challenge, they argue that "monitoring and logging will be enhanced with more interactive components for intermediate stages of active workflows." We did not find any work that registers steering actions in dynamic workflows in logs or provenance databases. Thus, there is no related work on tracking steering data and querying workflow data considering steering actions. Therefore, we initially analyze the main issues on computational steering, and then the related work on HPC computational steering, afterwards steering in application-specific scenarios, and finally we discuss steering support specifically in Parallel Workflow Management Systems (WMSs) and science gateways. The capability to track steering actions we are proposing is complementary in systems that already provide steering support.

Mattoso *et al.* [1] investigate six aspects of computational steering in large-scale workflows: interactive analysis, monitoring, human adaptation, notification, interface for interaction, and computing model. Despite the importance of each of them individually, the first three are essential for online analysis, tracking and evaluation steering actions (the challenges addressed in this work). Users will know how to adapt the workflow if they can analyze intermediate data during a long-term execution. Online provenance data management is an essential asset for interactive intermediate data analyses and monitoring, which are important ways to help gaining insights from the data being generated during execution. For monitoring, users set up monitoring analyses and wait for the results to be generated. Results might be presented as graphical dashboards or three-dimensional *in-situ* data visualizations. As users gain insights from monitoring results, new data exploration through interactive analysis can be done, and the monitoring can be adapted [15]. Human adaptation is the most important aspect of computational steering. There are several types of expertise in humans that are involved in a long-lasting workflow execution [15]. Domain scientists (*e.g.*, biologists, geologists) are experts in defining the hypothesis behind the experiment and interpretation of the results. Computational scientists (*e.g.*, bioinformaticians, numerical analysts) are experts in programming the computational models that do the simulations or in using programs that require HPC. They usually also have a good knowledge of the application domain. Computational scientists are the users responsible for computational steering [1–10].

Data-oriented solutions for workflows facilitate online human adaptation. When a user adapts the dataflow, new data (user steering data) are generated, and thus their provenance must be registered. Not tracking may negatively influence results reliability, validation, and reproducibility.

For example, if a user removes subsets of a dataset (data reduction), the tasks (execution data) that would consume them will not need to be processed [15]. If a user fine-tunes parameters of a program, the overall result may be changed. In addition to reliability and reproducibility, having such data enables users to learn from their own adaptations: for example, they may find that when they tune certain parameters to a given range of values, the convergence of the linear equation solver improves by a certain amount. Finally, these adaptation data allow for building AI-based systems that help users while they are steering simulations [24], as they can extend their training database with provenance of adaptations.

Computational steering in parallel applications is a necessity for HPC users for decades [6]. Several systems provide steering support. Examples are SCIRun [25], CUMULVS [26], ParaView Catalyst Live [21], GRASPARC [27], Cactus [28], RealityGrid [29], and many others [2,5,6,30–34]. Similarly to our approach, these systems also allow for monitoring and parameter tuning, and require code instrumentation via libraries or API calls, which is an approach often adopted in Computational Science and Engineering [2,22,32] applications. Nevertheless, tracking of steering actions is not provided in any of those systems.

In addition to those systems, there are steering solutions for specific applications or domains. Often, examples that need user steering come from parallel scientific applications in the Oil and Gas industry [35]. For instance, BSIT [12] is a platform tailored for seismic applications that supports adaptations in parameters, programs, datasets, but it does not register provenance or allow for adaptation data analysis. Other examples are applications for Computational Fluid Dynamics [36].

With respect to WMSs, only a few [7–10] support user steering. Pegasus [37] provide a database with execution data to help debugging. Lee et al. [7] propose an extension to Pegasus to execute scientific workflows adaptively based on the analysis of Pegasus' database. However, in case a replacement on a data transformation occurs, because the adaptation is not registered, analyzing the average of execution time of a data transformation might give inconsistent results. Likewise, OpenMole [8] is a WMS for simulation models that need continuous adaptation and improvement. Users can replace programs in the workflow during its execution. Due to the lack of registering when, what, and who did the replacement, a different user may choose an already tested configuration disregarding previous efforts. Both Pegasus with its extensions [7] and OpenMole could benefit from our approach to register their supported steering actions.

FireWorks is a WMS [9] that uses a DBMS-driven workflow execution engine. It has a JSON-based approach for state management and uses MongoDB to query JSON documents to monitor workflow execution. However, no other steering action is supported. Copernicus WMS [10] also allows for dynamic workflow steering via parameter tuning, sharing similar motivations to ours. It also aims at analyzing data to steer exploration towards undiscovered regions of a solution space. Typical parameters tuned by users are initial seeds, number of samples, and parameters specific to the analysis method. These systems evidence the need for tracking and querying steering actions like we propose in this work.

Chiron WMS enables users to change filter values and adapt loop conditions of iterative workflows [38], and reduce input datasets [15]. These works show that online adaptations significantly reduce overall execution time, since users can identify a satisfactory result before reaching the programmed number of iterations. However, tracking the adaptation has not been addressed in Chiron.

WorkWays [4] is a powerful science gateway that enables users to dynamically adapt the workflow by reducing the range of some parameters. It uses Nimrod/K as its underlying workflow engine, which is an extension of the Kepler workflow system [39]. It presents several tools for user interaction in human-in-the-loop workflows, such as graphic user interfaces, data visualization, and interoperability among others. Such graphical functionalities can highly benefit the user experience with the steering solution, and hence could be incorporated to DfAdapter for future work.

WINGS [40] is a WMS concerned with workflow composition and its semantics. WINGS facilitates the iterative process of designing workflows. This is complementary to our solution, as we need to identify the dataflow behind a computational model or simulation before the execution starts. WINGS also focuses on assisting users in automatic data discovery. It helps generating and executing multiple combinations of workflows based on user constraints, selecting appropriate input data, and eliminating workflows that are not viable. However, it does not allow for online parameter tuning, nor does it record the provenance of adaptations at runtime.

While WMSs and science gateways provide for efficient parallel workflow execution, this can be an issue when the workflow is already a parallel application. Simulations that use highly parallel libraries or adaptive parallel algorithms, already implement parallel execution control and scheduling on HPC machines. Often, this application parallelism conflicts with the scheduling and parallel execution of WMSs or science gateways.

Therefore, none of these systems, WMS, science gateways or other systems with steering support, provide steering action data tracking. The steering action definitions in Section 4 and the system design principles presented in Section 5 to track steering actions give directions that may complement current approaches to add the track of steering actions for dataflow analysis integrated with dynamic steering, following data provenance standards, all with negligible performance overhead.

## 4. User Steering Actions Definitions

Two data categories should be analyzed online to support human adaptation: *domain dataflow* and *workflow execution* [1]. While workflow execution is often associated to the control of task flow between chained activities [14], dataflows are often associated to datasets being transformed by the chaining of data transformations [16]. Each data transformation operates on input datasets and transforms it into output datasets. In parallel executions, elements in the datasets are mapped to workflow activity *tasks*.

**Domain dataflow.** The datasets that are produced or generated in the flow between data transformations are part of the domain application data that compose the domain dataflow. To analyze intermediate data with its context, domain dataflow must be available for interactive analysis and monitoring, while the workflow runs. Keeping track of the raw data files while keeping their context and relating their content to provenance improves online data analyses [16].

**Workflow execution data.** Data related to the workflow execution performance is very helpful in interactive data analysis, monitoring, and debugging. Users may monitor the execution data to control the amount of computational resources being used. Users are frequently interested in knowing how long tasks are taking or how much memory or CPU they are consuming. This information can deliver interesting insights when linked to domain dataflow data. For example, users can investigate which values are making a task consume more memory than expected.

Scientific workflows are data-centric and so are steering actions. Therefore, we follow a dataflow approach as opposed to a workflow control-based approach. Inspired by dataflow concepts proposed by Ikeda *et al.* [41], in previous works [15,16] we proposed a conceptualization for the flow of data elements in running workflows. In the present work, we refine and extend such concepts aiming to add semantics to data elements and to define steering actions. These semantics represent the role of data elements in the steered execution, for example, a parameter and a loop condition. Next, we define these concepts formally.

**Definition 1: Dataset, data elements, and data values.** A dataset $DS$ is composed of data elements, *i.e.*, $DS = \{e_1, \dots, e_m\}$. Each data element $e_i$, $1 \leq i \leq m$, is composed of data values, *i.e.*, $e_i = \{v_1, \dots, v_u\}$. Datasets are further specialized into Input Datasets ($I_{DS}$) and Output Datasets ($O_{DS}$).

**Definition 2: Data schema and attributes.** Data elements in a dataset $DS$ have a data schema $\boldsymbol{S}(DS) = \{a_1, \dots, a_u\}$, where each element data value $v_j$ has an attribute $a_j$, $1 \leq j \leq u$. Thus, an element data value can also be represented as a set of ordered pairs $\{(attribute, value)\}$, *s.t.*, $e_i = \{(a_1, v_1), \dots, (a_u, v_u)\}$. Moreover, attributes have a data type (*e.g.*, numeric, textual, array, etc.).

**Definition 3: Data transformation.** A data transformation is characterized by the consumption of one or more input data sets $I_{DS}$ and the production of one or more output data sets $O_{DS}$. A data transformation is represented by $DT$, where $O_{DS} = DT(I_{DS})$.

**Definition 4: Data dependency.** Let $DT_\alpha$ and $DT_\beta$ be data transformations and let $\{e\} \subset DS$ be a set of data elements produced in an output dataset $DS$ generated by $DT_\alpha$. If $DT_\beta$ consumes $\{e\}$, then $DS$ is also an input dataset of $DT_\beta$. In this case, there is a data dependency between $DT_\alpha$ and $DT_\beta$ through $\{e\} \subset DS$. A data dependency is represented as $\varphi = (\{e\}, DT_\alpha, DT_\beta)$.

**Definition 5: Dataflow.** A dataflow is represented by $Df = (T, S, \emptyset)$, where $T$ is the set of all data transformations participating in the dataflow, $S$ is the set of all datasets consumed or produced by the data transformations, and $\emptyset$ is the set of all data dependencies between the data transformations (adapted from background work [15,16,41]).

**Definition 6: Semantics of attributes.** We further group each attribute $a_i \in \boldsymbol{S}(DS)$ by its semantics $\Sigma(DS)$, so that: $\Sigma(I_{DS}) = \{F_I, V_I, P_I, L_I\}$ and $\Sigma(O_{DS}) = \{F_o, V_o, C_o, L_O\}$, where:

- $F_I$ and $F_O$ contain attributes that represent pointers to input and output files, respectively.
- $V_I$ and $V_O$ contain attributes for extracted data or metadata from input and output files, respectively.
- $P_I$ contains attributes for general purpose input parameter values of the data transformation.
- $L_I$ contains attributes used by in iteration loop, *i.e.*, used for data transformations that evaluate a loop.
- $L_O$ contains output values especially related to an iteration in case of data transformations that evaluate a loop.
- $C_O$ contains attributes for any output values that are explicit data transformation results.

Such added semantics improves the data modeling of the dataflow and allows specifying which attributes of a $DS$ are parameters to be steered. Parameters $P_I$ are the main target of fine tunings. For example, parameters are numerical solver configurations, thresholds, or any other parameter that can be adjusted.

$F_I$ and $F_O$ are often large raw (textual, imagery, matrices, binary data, etc.) scientific data

files in a wide variety of formats depending on the scientific domain (*e.g.*, FITS for astronomy, SEG-Y for seismic, NetCDF for computational fluid dynamics simulations). These data are typically not tuned, but are important for data analyses.

In the case of output data, examples are QoI. Some applications write calculated values, like the QoI results of a data transformation into files and they often need to be tracked. $V_O$ represents these special resulting extracted data, which are often scalars, useful for domain data analyses [15,16,18]. $V_I$ and $V_O$ can be seen as views over the actual large raw data files, as users can have a big picture of the content of the files through them.

Besides large scientific data files produced by data transformations, they may produce explicit output results, $C_O$, often scalar values or simple arrays that are very meaningful for the overall result. Since they may be of high interest for the user, these values are typical provenance data that need to be registered.

Moreover, the semantics of a dataset $DS$ may not be applicable to all attributes. For example, if a data transformation does not evaluate a loop, $\Sigma(DS)$ of this data transformation does not contain $L_I$ or $L_O$. Examples of $L_I$ are loop-stop conditions (*e.g.*, "*max*" in case of "*while counter* < *max"* loops or "*threshold*" in case of "*while error* > *threshold*" loops), or any other parameter used inside the iterations. In data transformations that evaluate loops, each iteration may be modeled as a loop evaluation execution and produces $L_O$. They are attribute values that contain current values being used to evaluate a loop, which are updated at each iteration (*e.g.* "*counter"* or "*error*").

**Definition 7: Steering action.** A steering action $SA$ is an interaction between a user who analyzes or monitors or dynamically adapts one or more elements of $D$:

$$D' \leftarrow SA_\alpha(D)$$

where $D$ is a $Df$ or a $DT$ or a $DS$ and $\alpha$ is a steering action clause that defines the analysis or monitoring or adaptation that result in $D'$.

For example, when $D$ is a $Df$, users might need to monitor (or analyze or adapt) the composition of data transformations of the dataflow. When $D$ is a $DT$, users might need to monitor (or analyze or adapt) the $DT$ structure. In case of $DS$, users might need to monitor (or analyze or adapt) data elements in the $DS$. Depending on the operand $D$, $\alpha$ specifies which elements the user will interact. When $SA$ is monitoring or analysis, $D'$ contains the result of the monitoring query or analysis. When $SA$ is adaptation, $D'$ contains the resulting data modified by the user.

**Definition 8: User steering data.** To register a steering action $SA$, user steering data need to be tracked. User steering data is denoted by $(D, \alpha, D', U, T)$, where $D, D'$ and $\alpha$ are the data and the clause, respectively, involved in $SA$; $U$ contains data about the user who performed $SA$; and $T$ is a set of data transformation executions related to $SA$. Any other data that benefit the register of the steering action $SA$ can optionally be tracked and associated to $SA$. For example, the current wall time at which the $SA$ occurred, or textual annotations informed by the user at the moment of $SA$ can benefit its register.

Considering that parameter fine tuning is the main action within adaptations in a $SA$, we define a special case of $SA$, named $Tune$.

**Definition 9: Tune**. $Tune$ is a steering action for parameter tuning as follows:

$$I'_{DS} \leftarrow Tune_{(\eta,C)}(I_{DS})$$

where the operand $I_{DS}$ contains old values of attributes being tuned into $I'_{DS}$ with the new values. $I'_{DS}$ follows the same schema $\boldsymbol{S}(I_{DS})$ and semantics $\Sigma(I_{DS})$. $(\eta, C)$ is the steering action clause. $\eta$ is a set of ordered pairs $(p, v)$, where $p \in P_I$ is the parameter being tuned and $v$ is its new value. $C$ expresses a predicate to address a specific data element that will have its parameters tuned. In case of an $I_{DS}$ that contains a single data element, $C$ is optional.

To register a $Tune$ operation, the user steering data tracked are: $(I_{DS}, \eta, C, D', U, T, d)$, where $d$ is an optional argument that contain useful data related to the steering action context.

## 5. DfAdapter

In this section, we present DfAdapter, a lightweight solution aimed at capturing provenance of online steering actions in dataflows and storing the related dataflow provenance to enable understanding of the impacts of the action. Section 5.1 shows the main system design principles followed by DfAdapter. Section 5.2 shows how steering actions are captured in different workflow execution models. Section 5.3 presents the system architecture. Section 5.4 provides a general overview of how to use DfAdapter. Sections 5.5 and 5.6 present the provenance data model and its implementation using the relational data model, respectively. Section 5.7 provides a formalism to calculate DfAdapter's overhead.

### *5.1 System Design Principles*

In this section, we explain the core system design principles followed by DfAdapter.

**Asynchronous service calls.** DfAdapter is coupled to adaptable applications, like systems that support computational steering [1,2] or adaptable simulations as our case study. In either case, an API for DfAdapter is used so it can be called from the adaptable application. Provenance capture calls are placed in monitoring points in the workflow code to capture provenance of the dataflow and execution of data transformations. Similarly, to capture provenance of steering actions, DfAdapter calls placed in the steering points, allowing DfAdapter to track the steering actions in data transformations at runtime.

Attaining low performance overhead is a basic requirement in DfAdapter, otherwise computational scientists, used to high performance systems, will not use the tool. For this, calls to DfAdapter are asynchronous, meaning that when the user adapts the running workflow, the track of steering actions is done in such a way that the main computational process will not wait until the track finishes. The same approach is valid for any added monitoring data tracking in the code. In addition, the most computationally costly components in DfAdapter, such as the ones that store steering data in the provenance database during workflow execution, are deployed in separate hardware, different from where the main computational process runs, but in same internal network (*e.g.*, the nodes in the cluster has local access to the node that runs DfAdapter's provenance server) to reduce communication costs, following *in-situ* and *in-transit* approaches [2]. This avoids making DfAdapter and the main computational process compete for resources. Following these principles, the utilization of DfAdapter attains low added performance overhead for provenance of steering actions, such as less than 1% in our case study (Section 7).

**Adapter service and communication between DfAdapter interface and the running workflow.** Adding steering points in an adaptable workflow means that in those points there will be data communication between the running workflow and DfAdapter, so that the data flowing in the workflow can be modified. To represent this communication between the front end (from which the user sends steering commands) and the back end (which receives the commands and effectuates an adaptation in the running workflow), we use the notion of an *adapter service*. The adapter service in an adaptable workflow has the communication protocol capable of adapting a running application. The basic idea is that the user uses DfAdapter interface that communicates with the front end of the adapter service, which sends steering commands to the back end of the adapter service that does the adaptation in the running workflow, and finally DfAdapter registers the provenance of the steering action. There are different ways to implement such data communication between the back and front ends of an adapter service [2,5,6]. DfAdapter can be coupled to any of these implementations. These implementations are the following.

*(i) File-based checks.* This is a simple yet widely used way to implement data communication [2]. In this case, there are files in a storage resource that are accessible both by the front and by the back ends of the adapter service. In that case, when the user uses DfAdapter interface to steer the workflow, the front end of the adapter service modifies a file according to the user inputs. When the program pointer in the running workflow reaches a steering point, the back end of the adapter service verifies if files were modified and, in case of modification, the adaptation is carried out and DfAdapter is called to register the adaptation. Although file-based checks are a simple approach, it is widely used especially by users that implement their own *ad-hoc* way to make their simulation steerable, as in our case study. However, it requires that front and back ends share access to files in a storage resource, which may not be always possible.

*(ii) Message passing.* It is another way to implement data communication. In this case, when the user uses DfAdapter interface, the adapter service's front end sends a message to its back end in the running workflow. When the steering point is achieved, the adapter service's back end verifies if a message has arrived and effectuates the adaptation accordingly, and DfAdapter is called to register the adaptation. MPI, sockets, or RESTful HTTP messages can be used as communication protocol to implement this. Many systems with steering support use message passing to implement data communication [29–31]. This is an alternative to file-based checks, as it does not require files to be shared in a storage resource by the adapter's front and back ends.

*(iii) DBMS-driven*. It is an alternative to message passing and file-based checks. It is similar to file-based checks in the sense that there is a DBMS that is accessible both by the DfAdapter interface (via the front end in the adapter service) and the running workflow (via the back end). It is similar to message passing in the sense that it does not require files to be shared in a storage resource. Rather, data that can be modified at runtime are managed by the DBMS that can even run in-memory, depending on the DBMS. In this implementation, when the user adapts using DfAdapter, the adapter front end modifies data in the DBMS. When the program pointer achieves the steering point in the running workflow, the steered end checks if the data have been modified, carries the adaptation accordingly, and DfAdapter is called to register the adaptation. We implemented the data communication and steering action tracking in a synthetic workflow example using the parallel frameworks Apache Spark and Redis, a lightweight in-memory Key Value store, as the DBMS between the workflow and DfAdapter. The source code is available on GitHub [42].

**DBMS and data model for the Provenance Database.** DfAdapter needs a DBMS to manage the provenance database. Several data models can be used for provenance databases, such as graph and relational data models. The usage pattern in DfAdapter is that the running workflow only produces insertions to the provenance database, while the user typically runs provenance queries for online data analyses to support decision-making, *i.e.*, OLAP queries. This usage pattern, both by the workflow system and by the user, is benefited from column-oriented relational DBMSs, as shown in some of our previous works [16,18]. Moreover, since there may be many appends to this database during execution, the DBMS must be able to handle parallel requests from clients. Thus, DfAdapter follows this principle and uses a DBMS, called MonetDB[†], which has these characteristics.

**Provenance data modeling.** Provenance data management is at the core of DfAdapter. Instead of creating new standards, DfAdapter follows the well-stablished W3C PROV recommendation, and extends to add the specific parts for the track of parameter tuning. By adhering to W3C PROV standard, DfAdapter aims at allowing for interoperability among provenance databases. In addition, another important principle in DfAdapter is that the provenance data model is abstract and flexible enough to be used in different domains or applications. Our previous works show that similar provenance data modeling used by DfAdapter has been shown useful to capture relevant domain-specific data as well as generic dataflow provenance in other applications, such as in the Oil and Gas domain [15] and Astronomy [16]. The provenance data model (Section 5.5) is thereafter implemented in a relational data model (Section 5.6).

### *5.2 Keeping Track of Parameter Tuning in Different Workflow Execution Models*

Workflow execution models are acyclic or cyclic. Acyclic model is the most commonly supported in workflows (often modeled as a Directed Acyclic Graph), although the cyclic model need to be supported for extreme-scale workflows [14]. We design our solution for tracking user steering actions to support both models.

In case of acyclic models, after the user tunes attributes in an $I_{DS}$, provenance data collectors register which attributes were modified with their old and new values, and execution data of tasks that were running at the time of the adaptation.

Both sequential and concurrent execution models can be iterated in a cyclic model. Thus, at runtime, when the workflow is running in a specific cycle, a user can tune parameters. Cyclic execution models can be further distinguished between (i) loops without dependencies between iterations, also known as *parameter sweeps*; and (ii) loops with dependencies. Examples of loops with dependencies are counting loops, such as *for i=0; i < max; i++*, or conditional loop, such as "*while error > threshold"*. Additionally, Dias *et al.* proposed "external steering" loops, where the user adapts loop-stop conditions [38].

For (i) parameter sweep loops, the user may want to modify parameters that are to be swept. For (ii) loops with dependencies, the current iteration counter is an important value to be tracked. In those cases, the evaluation of a loop can be modeled as a data transformation [38] in a dataflow. Thus, the $Tune$ operation represents the tuned attributes that will be used inside the cycle in $L_I$; and $d$ (optionally tracked when $Tune$ occurs) contain the current iteration counter. Thus, $d$ is tracked with the current iteration counter of the loop, alongside

[†] https://www.monetdb.org

with $\eta$, $\theta$, $\epsilon$, as in previous execution models.

### *5.3 DfAdapter Architecture and Details*

In this section, we present details about DfAdapter architecture following the design principles previously presented. DfAdapter controls the front end of the adapter service. That is, when the user submits a steering command using DfAdapter interface, it registers the beginning of the steering action and makes the front end of the adapter service call its back end. DfAdapter's API method to be inserted in the steering points of the workflow code implements the $Tune$ operator. When the back end of the adapter service effectuates the adaptation, an API call to DfAdapter is executed, which tracks the steering action and stores in the provenance database. Figure 3 shows DfAdapter system architecture and Figure 4 shows an UML sequence diagram that represents the steps that occur when the user issues a steering command.

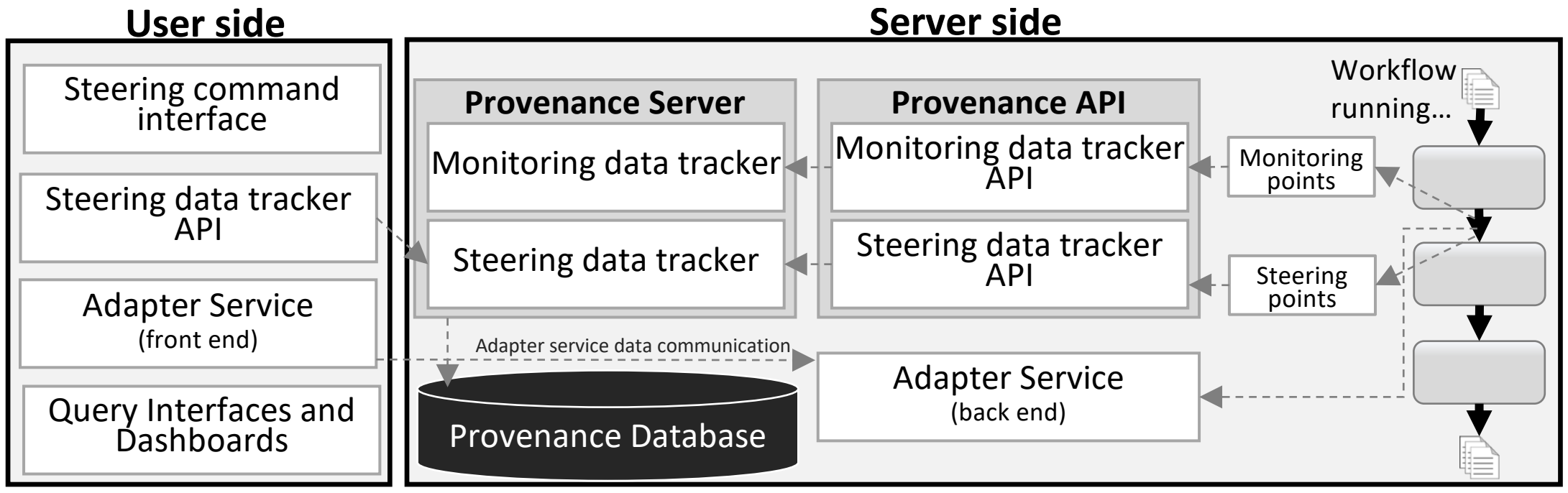

**Figure 3. DfAdapter system architecture.**

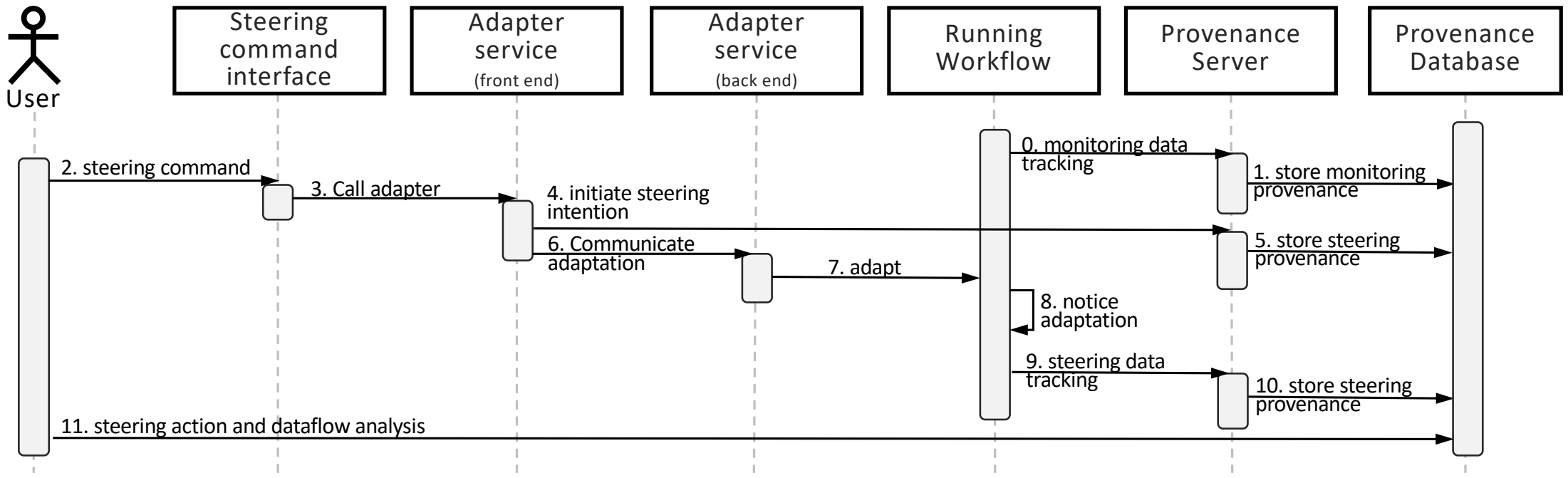

**Figure 4. Sequence diagram for the track of steering actions.**

The sequence of steps that occur when a user steers using DfAdapter are as follows: First, during workflow execution, (0) monitoring data specified in monitoring points are sent to the `Provenance Server` via `Monitoring data tracker API` calls. Then, (1) `Provenance Server` stores monitoring provenance in the `Provenance Database`. While the workflow runs, user can use `Query Interfaces and Dashboards` to follow the intermediate data results and decide for a steering action. If the user decides for a steering action, (2) the user sends a steering command using DfAdapter's `Steering Command Interface`, which (3) calls the `Adapter Service front end`, which (4) calls the `Steering data tracker API` method to (5) register the beginning of a steering intention. The `Adapter Service front end` reacts to DfAdapter's call and (6) communicates with the `Adapter Service back end`, which (7) has adapters that are able to effectuate the adaptation. When (8) the running workflow notices that an adaptation occurred (*e.g.*, it verifies that a file or a data structure, depending on the adapter service implementation, has been changed because of a steering action), the (9) `Steering data tracker API` method inserted in the steering point is called. (10) `Provenance Server` receives the calling and stores steering provenance in the `Provenance Database`. After that, the workflow continues to run normally together with monitoring data that are continuously tracked and stored, and (11) the user can run user steering action analysis.

The $Tune$ operator is implemented in DfAdapter system in the `Steering data tracker API` method inserted in the steering point (9th step in the sequence diagram Figure 4). It is implemented as shown in Algorithm 1, where we denote as $\theta$ the set of ordered pairs with the old values for the tuned parameters. Algorithm 1 is responsible to register new domain data that were modified in the adaptation as well as register their corresponding old values. It tracks current execution data, iteration counter values (in case of iterative workflows), and user data. Then, it stores user steering data relating to all other data being continuously tracked during workflow execution.

**Algorithm 1: DfAdapter using the *Tune* operator.**

**Input:**
$I_{DS}$: **The** $I_{DS}$ in **the dataflow containing the parameters to be tuned.**
$\eta$: **key-value data structure containing the parameters and their new values**.
$C$: **(optional) criteria to specify the data element that will be adapted.**

prov ← get_provenance_server() *//programming interface to the Provenance Server*
$\theta \leftarrow \varnothing$
$d \leftarrow \varnothing$
$T$ ← prov.get_running_execution_data()
$U$ ← prov.get_user()
$t$ ← get_current_wall_time()
current_data_element ← prov.get_element($I_{DS}, C$) *//if C is null, get the only data element in* $I_{DTD}$
attribute_semantics ← prov.get_semantics($I_{DS}$)
**for all key-value pairs** (p, current_value) **in** current_data_element **do**
    **if** p ∈ keys($\eta$) **then**
        $\theta$[p] ← current_value
        **if** p ∈ attribute_semantics[$L_I$] **and** $d = \varnothing$ **then** *//tuning a loop attribute. Get iteration data*
            $d$ ← prov.get_current_iteration_data($I_{DS}$)
        **end if**
    **end if**
**end for**
prov.store_steering_data($I_{DS}, \eta, C, U, T, d, t, \theta$)

## 5.4 *DfAdapter Utilization*

To describe how DfAdapter is used, we resort to the methodology listed in Figure 2. We explain how it can be added to dynamic workflows before execution and how it can be used to track steering actions.

**Before execution.** The user identifies a workflow by specifying parts of the code that can be modeled as data transformations and their datasets, and the data dependencies. Monitoring and steering points are added, as in Figure 1 of our case study. Then, DfAdapter API calls are inserted in the workflow code to capture provenance of the steering actions. Figure 5 shows an example using an excerpt of our case study workflow code. In Line 6, a method calls the DfAdapter API that implements Algorithm 1. The remainder provenance methods (Lines 3, 10, 13, 16) contain library calls inserted in the user code for monitoring so the user will know how to steer during execution.

```
...
for (unsigned int t_step = p.init_tstep; (t_step < p.n_time_steps) && (time < p.tmax); t_step++) {
    provenance.initTimeLoop();
    if ( parameters_modified() ) {
        p = reload_parameters();
        provenance.steeringTimeLoop();
    }
    ...
    for (unsigned int nonlinear_step = 0; nonlinear_step < p.max_nonlinear_steps; ++nonlinear_step) {
        provenance.initFluidSolver();
        flow_system.solve();
        ...
        provenance.finalizeFluidSolver();
    }
    ...
    provenance.finalizeTimeLoop();
}
...
```

**Figure 5. Excerpt of libMesh-sedimentation code with provenance and steering calls.**

**During execution.** DfAdapter wraps the front end of the adapter service. When users use DfAdapter interface to adapt the running workflow, the provenance of the steering actions are captured. The interface is command line-based, to be used in a terminal connected to the HPC machine where the workflow runs. In the command line, users only need to inform the input dataset $I_{DS}$ to be adapted, and a simple key-value data structure containing the parameters and their new values. For flexibility, the key-value data structure can be passed directly using the argument `--p` or write it in a file and pass its path as the argument. We add an optional argument `--reason` to allow users to annotate that specific steering action. Keeping the interface simple helps computational scientists to adhere to DfAdapter utilization. Figure 6 shows an example of DfAdapter's command line interface.

```
1.$> ./DfAdapter --user='Bob'
2.$> ./DfAdapter --tune
        --dataset='I_Iteration_Params'
        --p='{"max_linear_iterations":500}'
        --reason="checking how linear iterations affects
                 convergence"
3.$> echo '{
      "flow_initial_linear_solver_tolerance": 1.0e-6,
      "amr/c_fraction": 1.0e-5
     }' > new-values.json
4.$> ./DfAdapter --tune
         --dataset='I_Iteration_Params'
         --p='new-values.json'
```

**Figure 6. Command lines to call DfAdapter.**

### *5.5 W3C PROV for the Provenance of Parameter Tuning*

In this section, we propose a data provenance representation of parameter tunings. We build on our previous PROV-DfA [3], a representation for provenance of steering actions in dataflows, which is an extension of the W3C recommendation PROV-DM [43]. In this paper, we specialize PROV-DfA for parameter tuning in user-steered dataflows. In Figure 7, we use a class diagram to present the provenance data representation for parameter tuning. The classes in white background represent prospective provenance and in gray background represent retrospective provenance. The main added class is `ParameterTuning`. Parameter tunings at runtime are registered as retrospective provenance as they occurred while the workflow is in execution.

`ParameterTuning` represents provenance of a $Tune$ operation (*c.f.* Definition 9). It has two relationships (`WasInfluencedBy`) with `AttributeValue`. The first one is to relate to the new values of parameters being tuned. Values of parameters are modeled as `AttributeValue` (derived from the prospective entity `Attribute`), part of a `DataElement` of a `Dataset` (the $I_{DS}$ having its parameters tuned). Using W3C PROV relationships, we model the new attribute value of a parameter being tuned as a revision of (`WasRevisionOf`) the old parameter value, which is also an attribute value; hence the auto-relationship in `AttributeValue`. Thus, these relationships are for representing the new and old values for the parameters tuned, *i.e.*, $\eta$ and $\theta$. The second `WasInfluencedBy` relationship between `ParameterTuning` and `AttributeValue` is to relate the tuning with $d$ values, which are also modeled as `AttributeValues`.

To relate the `ParameterTuning` with $\epsilon$, we add the relationship `WasInfluencedBy` between `ParameterTuning` and `ExecuteDataTransformation`, which is the most representative class for workflow execution data (Section 4). For user data $U$, we relate `ParameterTuning` with `Person`, via the added `WasSteeredBy` relationship. We also create a new class, `Adapter`, which is a PROV `SoftwareAgent`, to store data about the program or service that can effectively adapt the dataset. We relate the tuning with the `Adapter` class via `WasAssociatedWith` to explicitly represent which `Adapter` call was used to tune the parameters. For example, one could use this relationship to store the arguments used by the service call to adapt the dataflow. Finally, `ParameterTuning` can be further extended for any other data that the user may find relevant, such as descriptions for the tuning or the criteria $C$ used to select the data element that will be tuned.

Therefore, with this W3C PROV-extended provenance data model, we can represent provenance of online parameter fine-tunings in dataflows steered by users. In the next section, we present one possible implementation of this provenance model using the relational data model.

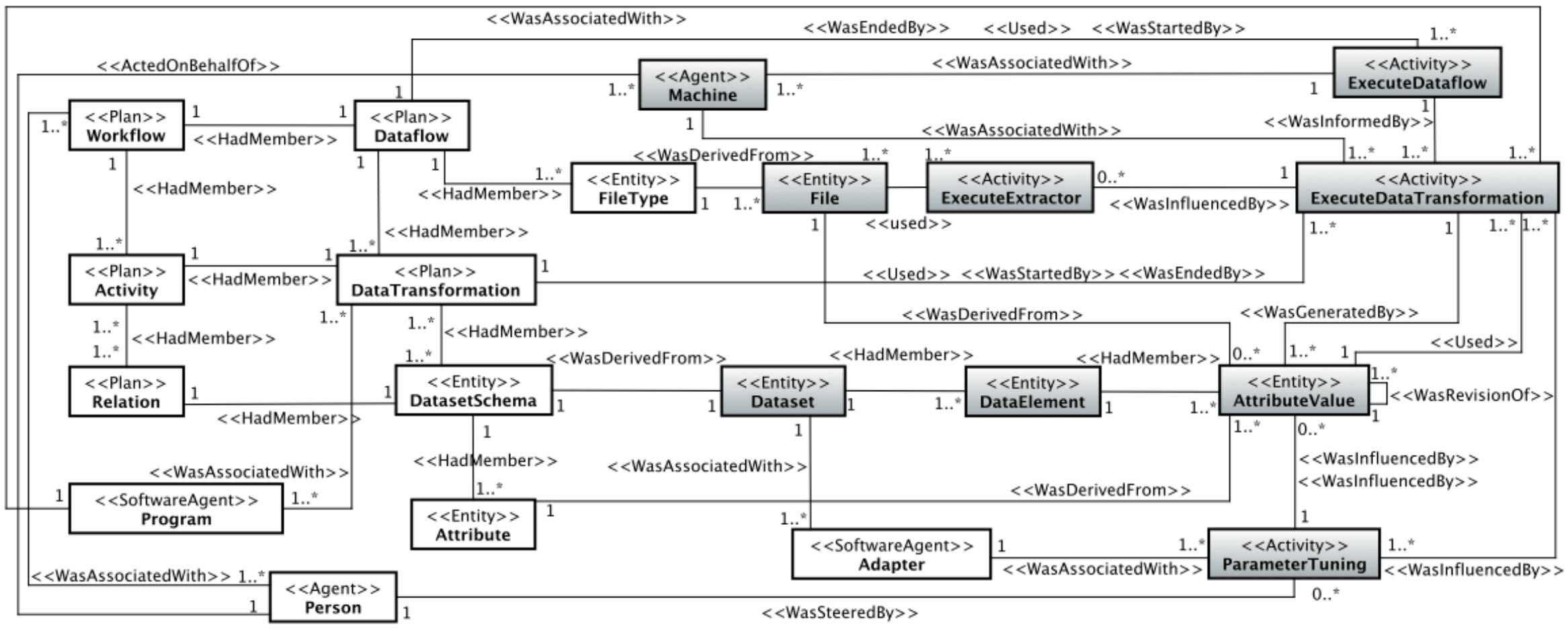

**Figure 7. PROV-DfA [3] with Parameter Tuning classes.**

### 5.6 *Implementing the Provenance Database Schema for DfAdapter*

We use the relational data model to represent the W3C PROV-extended provenance model presented in Section 5.5. An excerpt of the relational database schema is in Figure 8, whereas a complete figure can be found on GitHub [42]. Whenever a user issues a steering command to tune parameters, a new instance of parameter tuning action is stored in the `ParameterTuning` table. Since a parameter tuning may modify one or many attributes, and the same attribute may be modified by many steering actions, there is a many-to-many relationship between `ParameterTuning` and `Attribute` tables. The associative table, `ParameterTuned`, has fields to store old and new values. The $I_{DS}$ is a specialization of the table `Dataset`. Each tuple in `Dataset` table is a data element. Each `ParameterTuning` instance may directly affect one or many data elements in $I_{DS}$ and a same data element in $I_{DS}$ may be affected by many parameter tuning actions, hence there is a many-to-many relationship between `ParameterTuning` and `Dataset` tables, via the `ModifiedDataElement` associative table. Moreover, as $O_{DS}$ is also specialization of `Dataset`, we use `InfluencedDataElement` associative table between another many-to-many relationship between `ParameterTuning` and `Dataset` tables to store output data elements directly influenced by a tuning, such as iteration counter data in case of parameter tunings in data transformations that evaluate loops. Finally, we relate execution data about the current state of the execution when a tuning action happened via the associative table `InfluencedTask`. `Tasks` are directly mapped to `ExecuteDataTransformation` in the provenance model, and execution data are further extended with performance data via the relationship between `Task` and `Performance` tables. The person who steered and the adapter program used in that specific tuning are related and stored to `ParameterTuning`. Thus, because of these entities and relationships being populated during execution of the workflow in a user-accessible database, users can drive their analyses and decisions at runtime using these data.

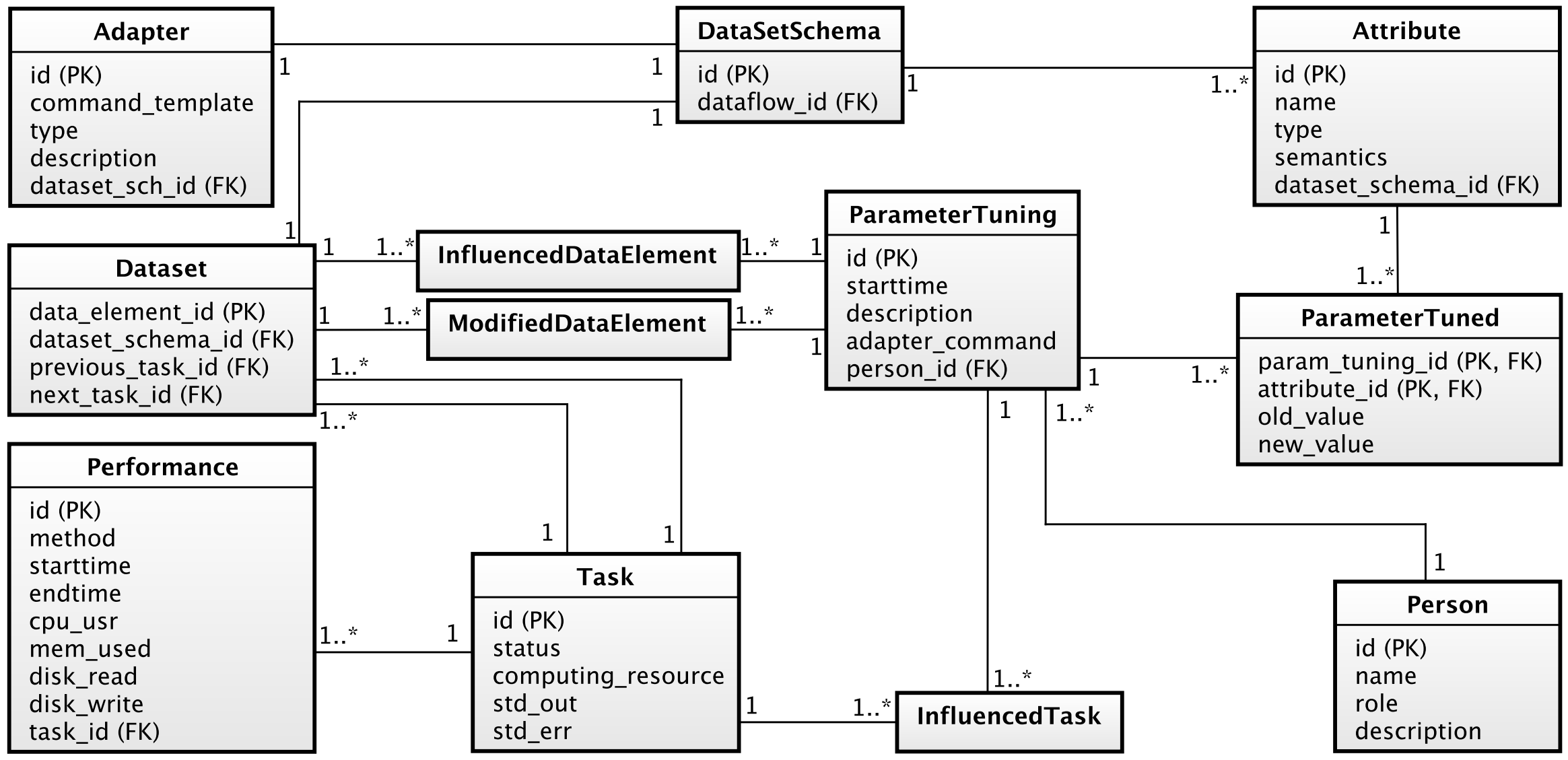

**Figure 8. Excerpt of the database schema.**

### 5.7 *DfAdapter Overhead Analysis*

The adoption of DfAdapter depends on how much execution overhead it implies. The overhead depends on data needed for monitoring and steering. For monitoring, it depends on the workflow data identified in the simulation code that needs to be tracked. That is, which input and output data values, for each data transformation, should be monitored during execution. For steering, which steering points should be added and how many steering actions actually happened during execution. In both cases, the overhead will depend on data collected for monitoring and for steering actions, always based on user decisions.

We use the dataflow concepts presented in Section 4 to express the overhead. Whenever a task $\tau$ is executed to perform a data transformation $DT_y$, the execution cost of $\tau$, $c(\tau)$, is given by its actual computational cost $comp(\tau)$ (*i.e.*, the inherent cost of executing $DT_y$) plus the introduced overhead $o(\tau)$. Let the overhead $o(\tau)$ of a task $\tau$ be expressed as a function of monitoring $m(\tau)$ and steering $s(\tau)$ overhead as in

$$o(\tau) = m(\tau) + s(\tau) \tag{1}$$

The overall overhead is given by the sum of $o(\tau)$ for all tasks $\tau$, of all data transformations $DT_y$ in $Df$. Next, we detail the monitoring and steering components.

**Analyzing monitoring overhead.** Monitoring overhead $m(\tau)$ is defined by the provenance data tracking overhead $prov(\tau)$ and raw data extractions $ext(\tau)$ during each data transformation execution identified by the user as relevant for monitoring, as in

$$m(\tau) = prov(\tau) + ext(\tau) \tag{2}$$

where $ext(\tau) = 0$ if there are no extracted data values in the execution of $\tau$.

Provenance tracking overhead $prov(\tau)$ depend on the number of data values of each data element tracked at a task execution $\tau$. Each execution $\tau$ of a data transformation $DT_y$ consumes input data elements in $I_y$ and produces output data elements in $O_y$. In DfAdapter, data elements are stored at once in the beginning (input data elements) and at the end (output data elements) of each task $\tau$. Provenance tracking overhead is due to preparing provenance tuples to be sent to the provenance database. Since provenance management services and the database system run in a separate computing resource and sending provenance data to be stored occurs asynchronously, provenance tracking overhead account only for preparing tuples to be sent. This represents a very low overhead, in the order of few milliseconds per task.

The raw data extraction overhead $ext(\tau)$ depends on the number of data values the user wants to extract from raw data files at each execution of a $DT_y$. Let $V_\tau$ be the set of all data values extracted when $\tau$ is executed. Each extracted data value $v_i \in V_\tau$ has an associated data attribute $a_i$ in $V_I$ or in $V_O$, depending on if $v_i$ is in a data element in $I_y$ or $O_y$, respectively – *c.f.* Definition 6. $ext(\tau)$ for each $\tau$ to execute a $DT_y$ is therefore given by the summation of costs to extract each $v_i \in V_\tau$:

$$ext(\tau) = \sum_{v_i \in V_\tau} ext(v_i) \tag{3}$$

The cost to extract a data value will depend on application-specific raw data extractors [16]. Extracting data values from raw data files to store in a provenance database for monitoring is done synchronously. Depending on the amount of data and how the raw data extractor is implemented, overhead may not be negligible, as we show in previous works [18].

**Analyzing steering overhead.** The steering overhead occur in data transformations that have a steering point. Steering overhead also depend on when a steering action happens. When a steering action happens, all those operations presented in the sequence diagram of Figure 4 are triggered. Let $S$ be the subset of all data transformations $DT_y$ in $Df$ that have steering points. For example, in our case study, the data transformation that evaluates the time loop has a steering point. Thus,

$$s(\tau) = s_{point}(\tau) + s_{action}(\tau) \tag{4}$$

where $s_{point}(\tau)$ is the overhead associated to adding steering points to $DT_y$, and $s_{action}(\tau)$ is the overhead associated to DfAdapter to compute that a steering action happened. $s_{action}(\tau) = 0$ if no steering action has been associated to the task $\tau$ and $s_{point}(\tau) = s_{action}(\tau) = 0$, $\forall DT_y \notin S$. The overhead $s_{point}(\tau)$ is a simple check to verify if a data structure has been modified during execution. Such simple verifications are nearly constant and milliseconds-long.

**Putting it all together.** The overall cost $c(Df)$ to compute a dataflow $Df$ is given by the sum of costs to compute the actual computation, $comp(Df)$, provenance tracking, $prov(Df)$, raw data extractions $ext(Df)$, steering points $s_{point}(Df)$, and steering actions $s_{action}(Df)$, that is,

$$\begin{aligned} c(Df) &= comp(Df) + o(Df) \\ &= comp(Df) + m(Df) + s(Df) \\ &= comp(Df) + prov(Df) + ext(Df) + s_{point}(Df) + s_{action}(Df) \end{aligned} \tag{5}$$

where $c(Df) = \sum_\tau c(\tau),\ for\ all\ tasks\ \tau, for\ all\ DT_y$ in $Df$. Analogously, all components of $c(Df)$ can be obtained by the summation of each individual component for all tasks. That is, $prov(Df) = \sum_\tau prov(\tau)$, $ext(Df) = \sum_\tau ext(\tau)$, and so on.

Therefore, the overall cost of a dataflow depends on the number of workflow tasks and the overall DfAdapter overhead depends on the number of tracked tasks and raw data extraction. We may consider that raw data extraction is a powerful support for data analysis, but not necessarily for fine-tuning, which may rely on provenance data monitoring. The raw data cost will depend on how much the user is willing to pay for data analysis. Therefore, we may separate the raw data extraction overhead from the remaining overhead costs.

We observe, on the scientific domain, that $\tau$ is often a complex task, where its $comp(\tau)$ takes at least a few seconds, but often minutes long [44]. By analyzing the individual elapsed time of the components, $prov, s_{point}, s_{action}$ of $o(\tau)$, we observe that, on average, they are close to constant and typically milliseconds-long. Therefore, we can assume that in scientific applications $comp(\tau) >> o(\tau)$, which leads to the negligible overhead of tracking user steering actions. In addition, because such operations occur asynchronously and in a different computing resource, the time for the individual components of $o(\tau)$ is "hidden" by the actual computation, which is significantly higher. This contributes to reduce the impact on the workflow execution performance.

If we consider $ext(Df)$, which depends on the user settings, it is still typically very much smaller than all raw data that is being generated and stored on files. As we show in our real case study, the overall $o(Df)$, including the costs for $ext(Df)$, is less than 2%, which is still negligible.

## 6. DfAdapter in Action: Montage and Sedimentation workflows

In this section, we present two real-world workflows modeled using the dataflow concepts and illustrated with steering actions. The parameter tuning cases are presented in increasingly order of complexity. First (Section 6.1), we illustrate the $Tune$ operation applied to Montage [45], a well-known workflow with a parameter sweep execution model. Montage exemplifies the applicability of our solution in a simple, yet typical case. Second (Section 6.2), we apply $Tune$ to our case study workflow, libMesh-sedimentation [18], showing the impact of fine-tuning in the performance.

### *6.1 Steering in Montage*

*Montage* [45] is a toolkit for assembling Flexible Image Transport System (FITS) files into custom mosaics, used for identifying potential objects of interest in the sky. It has been used for large-scale data analyses in the astronomy domain since 2005. Montage provides a service to build mosaics, according to common astronomy coordinate systems, arbitrary image sizes and rotations, and all World Coordinate System (WCS) map projections. It uses algorithms to maintain the calibration and positional fidelity of images to provide mosaics based on user-defined parameters of projection, coordinates, and spatial scale. It has independent modules for analyzing the geometry of images, and for creating and managing mosaics.

Before executing the workflow in the HPC machine, the user prepares the input data to be processed. Montage's `mArchiveList` module can be used for downloading FITS files, which are the inputs of this workflow. Each execution of `mArchiveList` has the input parameters: `survey` (represent source of the astronomy repository – possible values are 2MASS, DSS, etc.), `band` (the band or filter of the downloaded images – possible values are j, h, k, dss1, dss1b, etc.), `location` (name or coordinate of a mosaic region), and `width` and `height` (size of the area of interest, in degrees). These parameters represent regions in the sky and can be used to drive the analyses, as certain regions may be less or more likely to contain interesting celestial objects or, depending on these values, the assembled mosaic figure may have a better or worse resolution in a specific region of interest. See `mArchiveList` module[‡] for further details on each of these parameters. Furthermore, the output of a `mArchiveList` execution is a file containing a list of URLs of FITS files that can be downloaded. Then, we download each of those FITS files and compress them in a zip file. The input parameters used to execute `mArchiveList` are modeled as $P_I$ attributes in an $I_{DS}$ named `I_List_FITS`. The parameter values used in each `mArchiveList` execution, to download a list of FITS files (compressed in a zip file) compose a data element in `I_List_FITS`. Thus, the parameter values in one data element in `I_List_FITS` identify one zip file. Figure 9 shows a small subset of `I_List_FITS`.

[‡] http://montage.ipac.caltech.edu/docs/mArchiveList.html

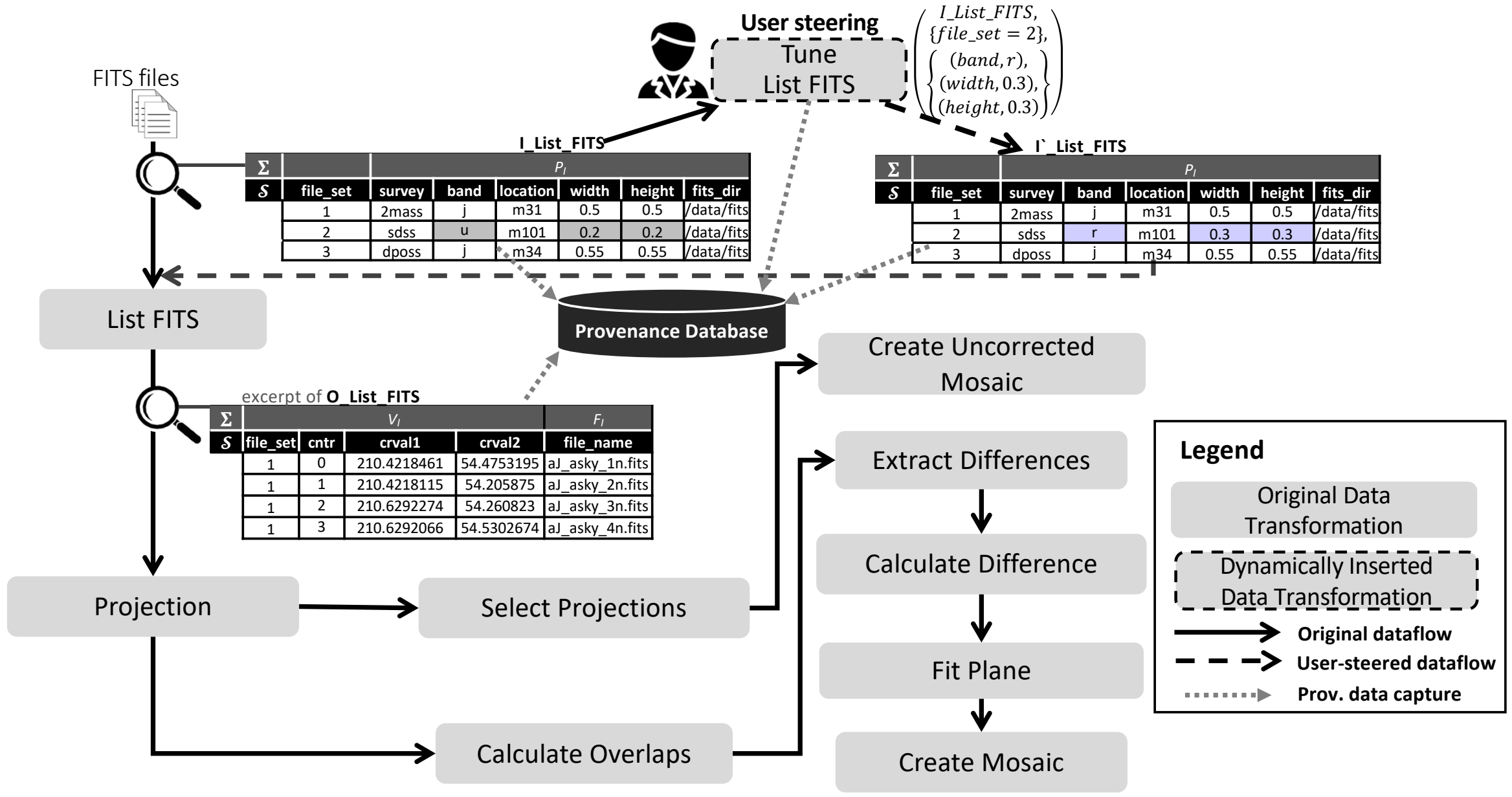

**Figure 9. User steering the dataflow in Montage workflow.**

Then, data transformations (mapped to Montage modules) in this dataflow are modeled as follows. The first data transformation (`List FITS`) extracts each of those zip files. Each input FITS file has 20 types of domain-specific values (modeled as $V_I$). The second data transformation (`Projection`) computes the projection of these astronomy-positioning references into a specific plane (extraction of 21 $V_O$ attributes). After that, `Select Projections` joins FITS projection files that are associated to the same mosaic (two $F_O$ attributes). `Create Uncorrected Mosaic` creates a mosaic without overlap interferences and color corrections and, as a result, it creates a JPG image (one $F_O$ attribute, the JPG file). The other data transformations from the Montage dataflow are defined to consider overlap interferences and color corrections to create a corrected custom mosaic.

Furthermore, Figure 9 gives details of the first data transformation, `List FITS`. `List FITS` data transformation uses the values in the data elements of `I_List_FITS` to get the zip file (stored in the directory informed in `fits_dir` attribute) to be processed in that execution. The workflow can be executed in a parameter sweep fashion (cyclic) with acyclic concurrent tasks, where the concurrent execution of each data element in `I_List_FITS` (going from this first data transformation until the last data transformation) represents a cycle in the parameter sweep (with attributes in `I_List_FITS` as the parameters to be swept). Then, for each FITS file in this extracted zip file, the data transformation `List FITS` creates a new data element in the output dataset (named `O_List_FITS`, which is input for the `Projection` data transformation). Each data element contains the file set it came from (`FILE_SET`) and a FITS file identifier (`CNTR`), allowing for tracing back, and two extracted elements from the input FITS files (`CRVAL1` and `CRVAL2`, modeled as $V_I$ attributes) that represent two coordinate values to determine a position in the native image coordinate system (*e.g.,* RA, Dec), and the FITS file (modeled as $F_I$ attribute for `Projection` data transformation). Other attributes are extracted at runtime in these data transformation and in all others throughout the dataflow, allowing for more online data analyses.

With such online data analytical support, the user is able to better understand the status of the execution and steer it. Users analyze the generated output mosaic images to investigate for interesting celestial objects in each analyzed region. Leaving the workflow to process the entire `I_List_FITS` dataset with no steering actions takes a long time, even though not all parameter values need to be processed. As the users gain insights from the online data analyses, they verify that certain values of the parameters in `I_List_FITS` (*e.g.*, `width`, `height`, `survey`) will not lead to finding interesting celestial objects. A bad choice for those parameters result in specific regions in the mosaic image with bad resolution or quality. If that region had an interesting object, it would be hard to identify. Thus, the user needs to tune the input parameters that are to be processed in order to change the region of interest. The execution may last for long hours. The user may decide on what is considered of interest several times. The situation may get worse when they tuned parameters that identify a region (with the objective of improving the quality of the resulting image), but the resulting mosaic does not have the expected quality or image resolution to validate her scientific hypothesis (the existence of a specific celestial object). Then, they need to tune the parameters in that region again to try to get better quality or resolution. Tracking such steering actions facilitates this process, allows for online data analyses of the steering actions, and improves reproducibility. In Figure 9, the user steers the dataflow during execution by tuning the parameters `band`, `width`, and `height` to specify a different file set to try to obtain mosaics with different resolutions in the region of interest. The $Tune$ operation transforms `I_List_FITS` into `` I`_List_FITS `` with the modified parameters, and the dataflow continues normally as if no tune happened.

### *6.2 Steering a Sedimentation Simulation*

libMesh-sedimentation workflow is our motivating case study (Section 2). Users typically set-up the QoIs and several parameters for the numerical methods. Examples of parameters are tolerances for linear and nonlinear solvers, number of levels for mesh adaptation, tolerances for space and time error estimates, etc.

Using the dataflow concepts, the QoIs in libMesh-sedimentation and the numerical solvers' parameters are modeled as data elements in datasets flowing in the dataflow. Moreover, function calls and other parts of the simulation source code are identified as data transformations. The dataflow has two acyclic setup data transformations, then the simulation enters in a time loop, configuring a cyclic execution model with loops with dependencies. There are five data transformations in the loop, including the solvers. Each solver runs in parallel, using all computing resources available in the HPC machine. The dataflow is modeled so that at each data transformation execution that evaluates the time loop, the parameters may be modified as the user steers. In Figure 10, we show the dataflow with an excerpt of the $I_{DS}$ (named `I_Iteration_Params` in the dataflow) that contains input parameters used inside the loops (i.e., $L_I$ attributes). Also, at each iteration, the `t_step` and `time` are captured, which are $L_O$ attributes in the `O_Iteration_Params`, which is the $O_{DS}$ for the time loop. Although only the datasets for the time loop are magnified in the figure, each arrow representing the original dataflow represents datasets for the data transformations, thus their data elements are being captured and stored in the provenance database.

In this scenario, we show the $Tune$ operator being used while the user steers (dashed lines in the figure) the dataflow by changing parameter values online. We can see that the original dataflow is modified by the user when the old value for the flow solver linear tolerance was tuned from $10^{-8}$ to $10^{-6}$, generating `` I`_Iteration_Params ``. When the user requests an adaptation, the $Tune$ operation will trigger the adapter to carry the adaptation and collect and relate steering data in the provenance database. In this case, as an iterative execution model with loops with dependencies, $Tune$ also relates the tuning with values of attributes in $L_O$ of the $O_{DS}$ of this loop evaluation data transformation.

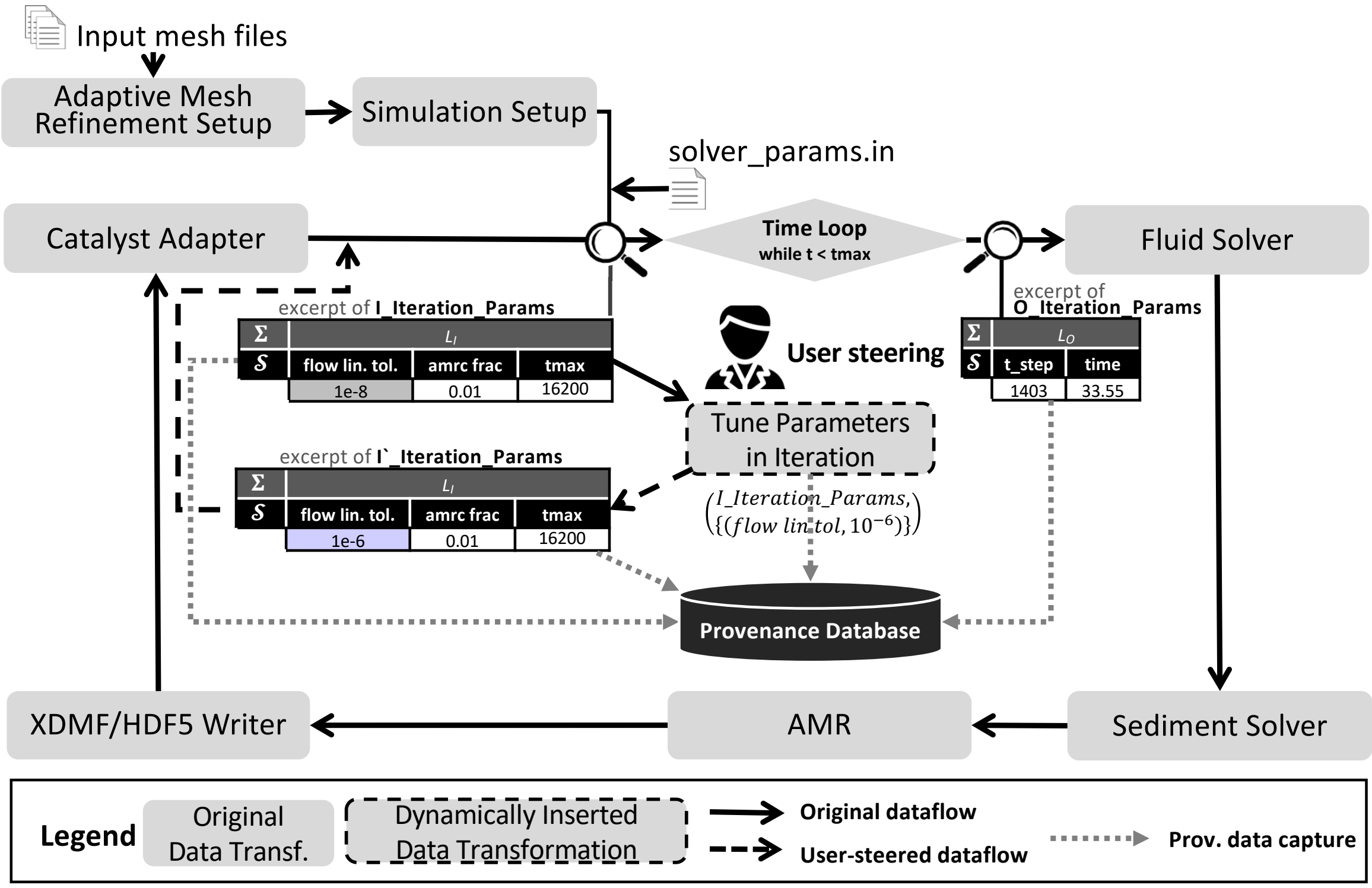

**Figure 10. User steering the dataflow in sedimentation simulation.**

## 7. Experimental Analysis

This section presents the tracking of online parameter fine-tunings in a real workflow from the Oil & Gas industry. We show that keeping a structured history of the steering actions supports the interpretability and validation of the results (Challenge 2). Also, we introduce how users can evaluate, at runtime, the impact of adaptations, through adaptation-aware online data analysis relating to provenance, domain, and execution data (Challenge 3). In Section 7.1, we present details of using DfAdapter in the case study and the experimental setup. In Section 7.2, we present a small-scale experiment from the same domain, highlighting different uses of our solution, and then a large-scale experiment in Section 7.3.

### *7.1 Implementation Details in a Numerical Solver and Experimental Setup*

**Implementation in a Numerical Solver.** We conduct the experimental evaluation on libMesh-sedimentation workflow, shown in Figure 1, which provides a real and rich case for parameter tuning. First, it is an HPC simulation with over 70 parameters, which may be modified by the user for better performance and accuracy of results [18]. Second, as this simulation may last for weeks, the user does several tunings and there is no tracking for them. Third, there is a strong potential for richer online data analyses with user steering data by correlating the steering data to domain-specific values (mainly QoIs) and other data in the provenance database. libMesh-sedimentation is implemented in C++ and its code with instrumentation for analysis and steering is available on GitHub [19].

The first step to use DfAdapter is modeling libMesh-sedimentation simulation as a workflow and identifying monitoring and steering points. Application-specific data are modeled as new tables of the relational database schema for the provenance database, and related to the existing ones accordingly (Figure 8). The main $I_{DS}$ that the user adapts is the input for the loop evaluation data transformation, named `I_Iteration_Params`, which contains input parameters for the numerical solvers. The users specify parameters in a setup configuration file. The simulation code checks, at every time step, if any modification has been made to this file. If a modification occurred, the parameters are redefined according to the new values. That is, libMesh-sedimentation workflow implements a file-based checks approach for the adapter service (Section 5.1). Modifications in this file are implemented as an adapter service front end, which basically receives parameters and new values, and modifies the file according to the inputs. Then, its execution is controlled by DfAdapter interface. The last step is to insert DfAdapter API calls in the steering points. In libMesh-sedimentation code, it is inserted immediately after the parameters are reloaded when there is a modification in the configuration file. Finally, when the user steers using DfAdapter interface, it captures provenance, domain and steering data every time it detects online user steering actions.

**Experimental Setup.** For the large-scale test, we use 480 cores from Lobo Carneiro cluster, an SGI ICE X with 252 nodes, each with a 24-core processor and 64 GB RAM, summing 6,048 cores and 16 TB RAM. The nodes are interconnected via FDR InfiniBand and share a Lustre file system with 500 TB. In this experiment, the provenance server and MonetDB are deployed in a separate node in the cluster, different from the ones used by the main computational process for libMesh-sedimentation. For the small-scale test, we use a Dell precision T3610 workstation, 8 cores, 16 GB RAM.

### *7.2 Small-scale experiment*

The small-scale experiment is used by scientists as a benchmark to evaluate sedimentation solvers. It simulates the laboratory test carried out by de Rooij and Dalziel [46] with a lock-exchange configuration. The objective of this experiment is to show the data analytical potential of our solution, how we record structured parameter-tunings, and how users can query the user steering data to enhance their analyses.

The computational setup used in this test case consists of a plane channel with dimensions 20 x 2 filled with sediments in suspension and clear fluid at rest. In the laboratory, a lock-gate is used to separate the fluids before the beginning of the experiment. When the gate is removed, a mutual intrusion flow develops in which the particle-laden front travels along the bottom to the right. In this simulation, the lock-gate is located at x = 0.75. The non-dimensional parameters used are Grashof number = $5.0\times10^{-6}$, Schmidt number = 1.0, and settling velocity 0.02. Adaptive mesh refinement is used to track the interface between sediments concentration and clear water. Figure 11 shows the concentration of sediments in suspension and the adapted mesh at simulation time t = 10.

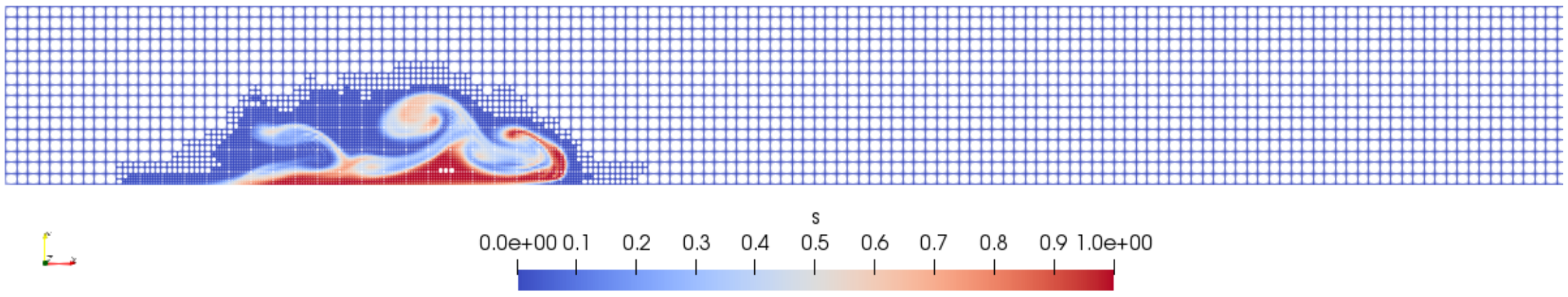

**Figure 11. 2D visualization of the tank and the concentration of sediments. This figure was generated at simulation time t = 10 using ParaView.**

In this simulation, the user is interested in analyzing possible performance gains when the number of nonlinear and linear (in this case, GMRES) iterations is tuned at runtime. Specific fine-tunings on different input parameters may impact the solvers and hence the simulation time considerably. During the execution, the user submits analytical queries to DfAdapter, addressing Challenge 1. Based on the analyses of nonlinear and GMRES iterations, the user decides to fine-tune the solver's parameters. In total, the user chooses to do six fine-tunings in 10 hours of simulation. Query 1 (whose description and tabular results are in Figure 12) shows the provenance of the adaptation. It lists all the parameters tuned by a user (say, Bob), correlated to the time steps. By using Query 1, other researchers are aware that Bob adapted this workflow execution six times. The times and values are well-structured and recorded in the provenance database by DfAdapter, thus addressing Challenge 2.

**Query 1:** List all user tunings correlating with time step.

This query does a join on tables: `ParameterTuning`, `ParameterTuned`, `InfluencedDataElement`, and `Attribute`, filtering by tunings made by 'Bob'. The result is:

| Parameter Tuning | t_step | Parameter Tuned | Old Val | New Val |
|---|---|---|---|---|
| 1 | 1401 | flow_initial_linear_solver_tolerance | 1e-8 | 1e-6 |
| 2 | 1474 | minimum_linear_solver_tolerance | 1e-8 | 1e-6 |
| 3 | 1484 | flow_initial_linear_solver_tolerance | 1e-6 | 1e-4 |
| 4 | 1755 | max_linear_iterations | 500 | 300 |
| 5 | 10061 | amr/c_fraction | 0.01 | 0.05 |
| 6 | 10128 | max_linear_iterations | 300 | 200 |

**Figure 12. Query 1 results.**

To inspect the consequences of adaptations, a more sophisticated analytical query is needed. Query 2 (whose description and tabular results are in Figure 13) shows the average values of strategic quantities ten iterations before and after each of the four fine-tunings. The results include nonlinear and linear (GMRES) iterations, which are output values of the solver, and the number of finite elements, which is an output of the mesh refinement process and depends on other inputs of the solver. This query shows an integration of provenance of the domain dataflow, performance data (average of elapsed times in 10 iterations), and the new fine-tuning data introduced in this paper. The results of Query 2 (we highlight the main findings) show that the Tunes #3, #4, and #6 impacted the average elapsed time and the average number of GMRES iterations, which are of high interest to the user. Tune #5 barely changed the other values but reduced the number of mesh elements by about 11.15%, while keeping the overall simulation accuracy. This reduction is important because when there are too many elements, out-of-memory errors may happen (see next experiment). In Figure 14, we plot the evolution of these variables over time and annotate the tunings (Tune #1 to Tune #6) so the user can evaluate the adaptations, addressing Challenge 3.

**Query 2:** Average of domain values (QoIs) and simulation time estimate 10 iterations before and after the tunings.

This query does a join on tables `ParameterTuning`, `ParameterTuned`, `Attribute`, `InfluencedDataElement`, `O_Sedimentation_Solver`, `O_Fluid_Solver`, `Task`, and `Performance`. It also does an average on the output values of `O_Sedimentation_Solver` and `O_Fluid_Solver`, and on `Execution Time` in `Performance` table. The result is:

| Parameter Tuning | Avg Time (s) Bef | Avg Time (s) Aft | Avg nonlin. Bef | Avg nonlin Aft | Avg gmres Bef | Avg gmres Aft | Avg Elems Bef | Avg Elems Aft |
|---|---|---|---|---|---|---|---|---|
| 1 | 17.3 | 18.5 | 3.8 | 3.9 | 2.03e3 | 2e3 | 5.32e3 | 5.38e3 |
| 2 | 16.9 | 18.1 | 4.1 | 4.3 | 2.05e3 | 2.03e3 | 5.44e3 | 5.41e3 |
| 3 | 17.4 | 13.2 | 4.2 | 4.3 | 2.02e3 | 1.54e3 | 5.45e3 | 5.43e3 |
| 4 | 12.7 | 9.6 | 3.9 | 4.2 | 1.49e3 | 1.01e3 | 5.51e3 | 5.49e3 |
| 5 | 14.4 | 14.8 | 4.3 | 4.0 | 1.06e3 | 1.01e3 | 6.28e3 | 5.58e3 |
| 6 | 15.6 | 11.2 | 4.05 | 4.1 | 647 | 445 | 5.72e3 | 5.62e3 |

**Figure 13. Query 2 results.**

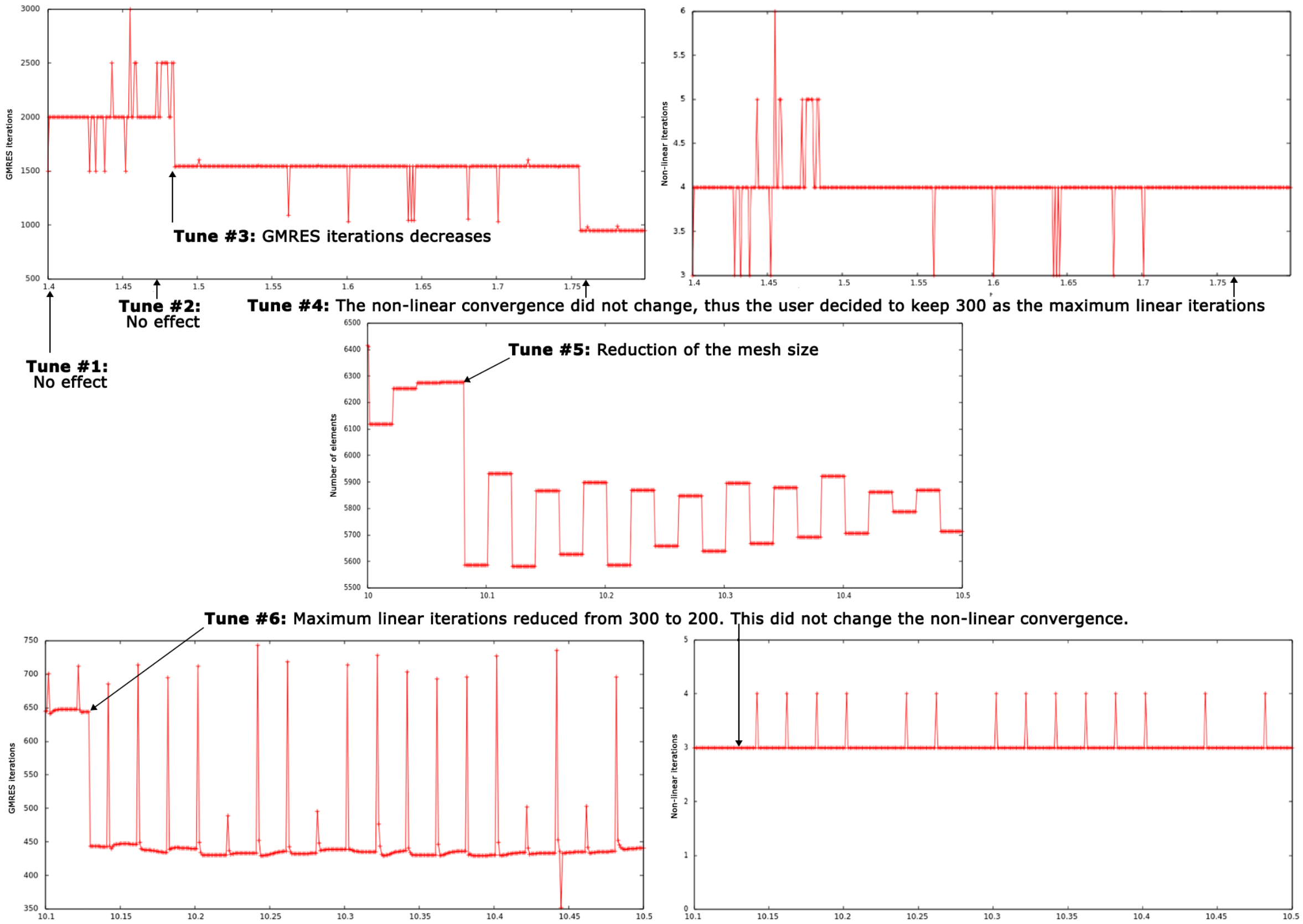

**Figure 14. Plots of monitoring queries for number of GMRES iterations, non-linear iterations, and mesh elements over time. We highlight the tune actions.**

Based on the adaptation-aware online data analyses, the user can evaluate decide whether or not new tunes are needed, also supporting the Challenge 3. Moreover, suppose a scenario where another research team analyzes the provenance of the results. The team sees abrupt changes in the results and can correlate these results with Bob's adaptation through SQL queries in the provenance database. They can check if sudden changes are related to one of the adaptations Bob did. Thus, they will have a better understanding of the results, thus addressing Challenges 2 and 3.

### *7.3 Large-scale experiment*

In this experiment, the user sets up the libMesh-sedimentation workflow with a simulation of the deposition of sediments carried by a turbidity current over a real experimental channel. A mixture of sediments is continuously injected into a channel that deposits sediments in the tank. The tank has length = 135, width = 40, and height = 50 (dimensionless units).

The dimensionless simulation parameters are settling velocity = $5.36\times10^{-6}$, Grashof number = $3.42\times10^{7}$, Schmidt number = 1.0, and fixed time step = 0.01. It uses a 3D simulation with a spatial discretization using an initial unstructured mesh with 1.2 million tetrahedra. AMR/C is employed and three levels of uniform refinement are applied before the time loop. The user specifies input parameter values for the sedimentation solver (i.e., linear and non-linear tolerances, maximum number of linear iterations, tolerances for AMR/C error estimation and refinement and coarsening fractions) aiming at attaining a high-fidelity simulation. One strategic simulation data that quantifies such level of detail is the number of elements obtained in the mesh refinement data transformation (second one in the time loop). Although a large number of elements in the mesh means high level of details, it also means more memory and time consumed by the simulation. Depending on the parameter values specified for the solver, the simulation might run out of memory. Thus, the user does not know beforehand which range of parameters is best for a good level of detail with acceptable memory consumption.

To support the user in following the evolution of strategic values, we use the approach presented in [15] to monitor results using provenance and domain data. We set up several monitoring queries to plot simulation data at each time step. One query shows linear and nonlinear iterations, residual norms, and the number of elements in the mesh at each time step. Additionally, ParaView Catalyst is set up to plot 3D visualizations of the channel and the sediment deposits over time. Then, the user sees, for example, that the number of elements generated by the AMR/C is close to a maximum preset number of elements. At that rate, the simulation may crash, running out of memory. The user knows that by changing some of the solver parameters, the number of elements tend to decrease. Thus, the user issues a command to adapt the solver parameters and DfAdapter automatically tracks and registers this tuning.

In Figure 15, we show the plot of the monitoring query for the number of elements. We see how the number is increasing when the user decided to fine-tune the input parameters online aiming at reducing the number of elements. This action prevented the simulation to result in an out-of-memory error, which would interrupt the simulation, requiring offline tunings and job resubmission to the HPC machine.

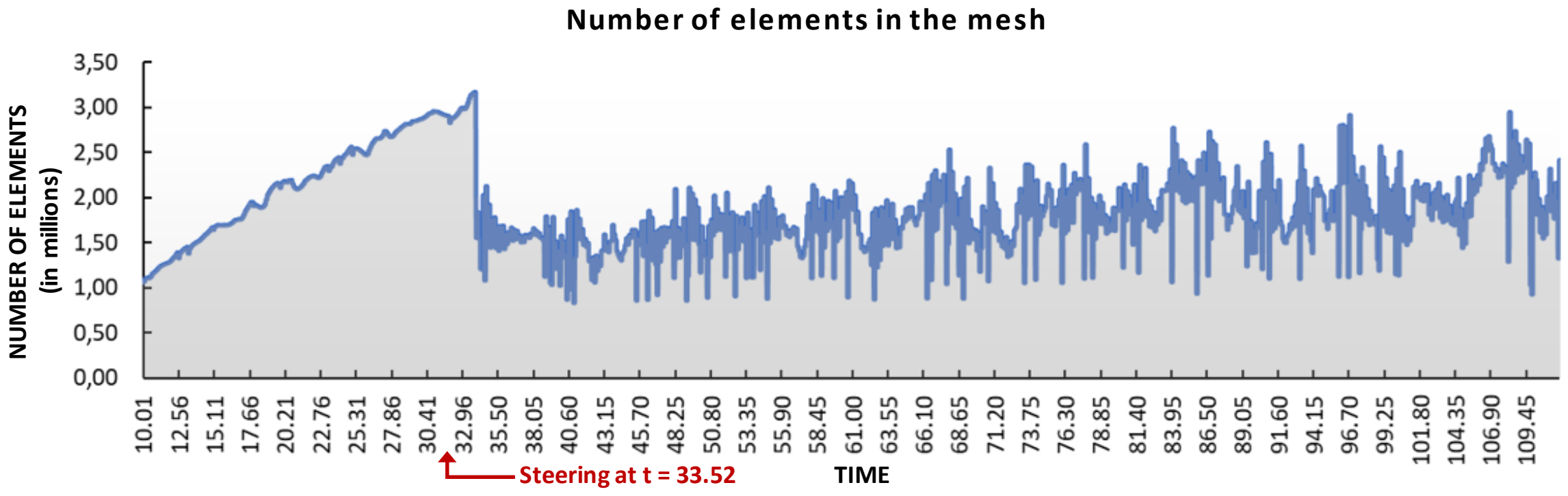

**Figure 15. Plot of monitoring query showing number of elements over time.**

In Figure 16, we show the 3D visualizations and the evolution of the strategic values and how the sediments flow in the channel over time. Then, the user can run analytical queries to analyze the consequences of the fine-tunings, like Queries 1 and 2. In Table 1, we show a small excerpt of these results, where we can see that the simulation time is cut down to 17 days, thanks to the fine-tunings. If we consider the average solver time by iteration before the fine-tunings, the simulation time would be approximately 27 days, *i.e.*, a reduction of 37%.

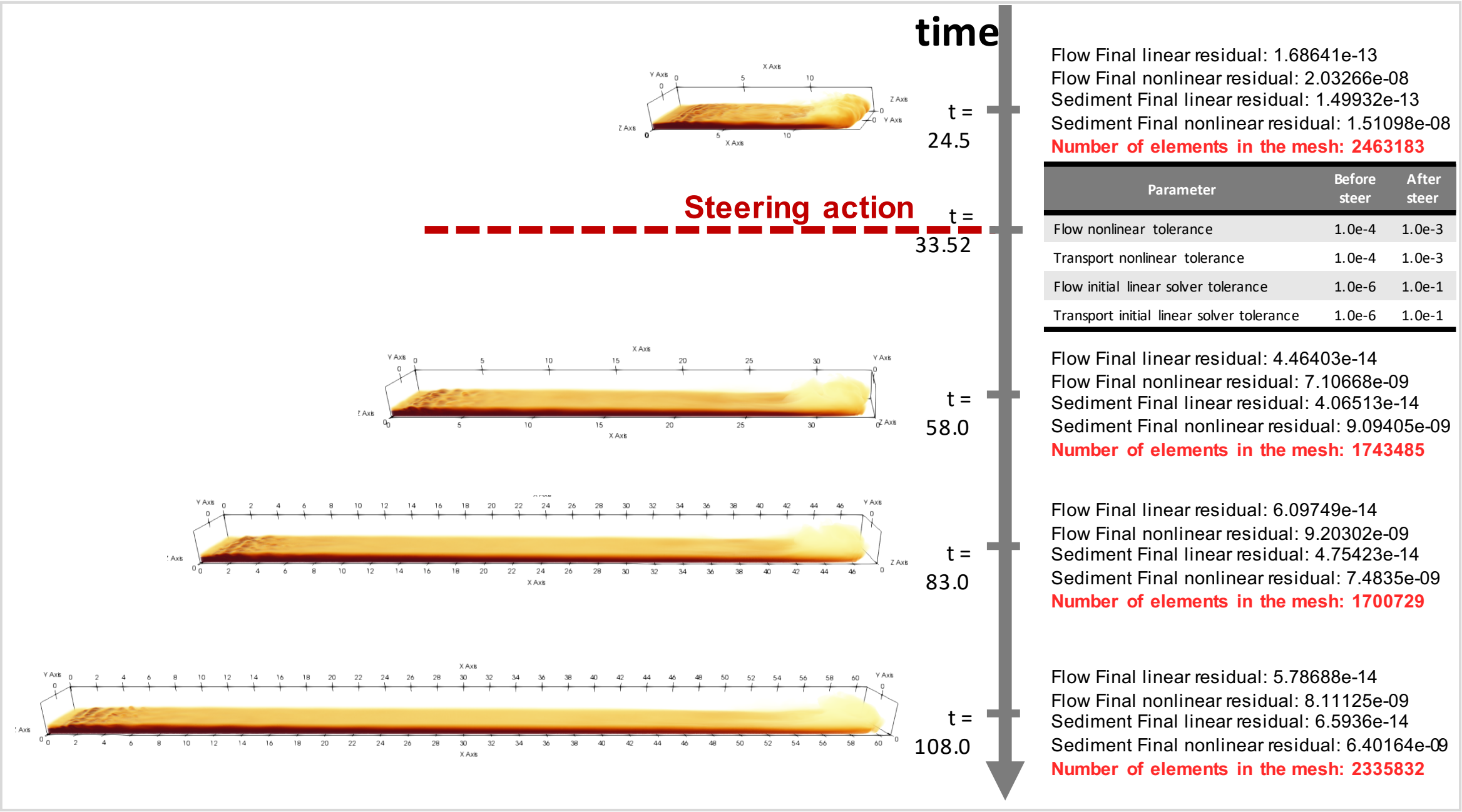

**Figure 16. Snapshots of 3D visualization of the tanks and the sediments over time. Steering action occurs at t = 33.53 and steering data are recorded.**

**Table 1. Results of parameter-tuning.**

| | Before | After | Reduction |
|---|---|---|---|
| Avg. Solver Time by iteration | 3.82 min | 2.21 min | 42.14% |
| Avg. Number of elements | $2.4x10^6$ | $1.7x10^6$ | 29.24% |
| Total execution time | (expected) ~27 days | (real) ~17 days | 37% |

**Analyzing the computational time in details.** We use the concepts and equations presented in Section 5.7 to analyze computational time of libMesh-sedimentation in this experiment and added overhead due to provenance capture, data extraction, and steering capabilities. Results are in Table 2. To obtain them, we first calculate each overhead component per task applying the Equations 1 to 4 using DfAdapter's logging data joining with tasks' performance data in the provenance database. Finally, we sum each contribution to the overall computational time as in Equation 5.

For monitoring, provenance-tracking overhead account for 0.3% caused by preparing the tuples to be sent to the provenance server. libMesh-sedimentation workflow has a steering point in the beginning of the time loop iteration. Raw data extractors extract convergence values from raw data files written as XDMF/HDF5 so the user can monitor and detect possible misbehavior of nonlinear and linear solvers. In total, these raw data extractions account for 1.49% of the total computation time. For steering, since libMesh-sedimentation uses a file-based checks implementation, it verifies if a file has been modified at each new time iteration. This file verification is synchronous because the simulation code must verify if a change has happened before it can continue. In total, this check at each new iteration adds 0.03% overhead. When a steering action happens, the internal data structure that contains the solver parameters is reloaded and steering data are tracked and sent to the provenance database. Since the user steered 6 times during execution of this workflow, the overhead for steering action tracking is close to 0%.

Such reduced overhead is due to our system design principles related to asynchronicity and to the fact that the most costly data tracking operations occur in a separate node. Also, because libMesh-sedimentation tasks are seconds-long on average (Figure 13), the distributed CPUs spend significantly more time computing the application tasks than computing provenance or steering functions. Therefore, considering approximately 17 days (about $1.4\times10^6$ s) of total execution time, provenance and steering tracking together account for less than 1% overhead, whereas summing with raw data extractions, the total overhead is less than 2%.

Table 2. Provenance and steering overhead account for less than 1%, whereas data extraction account for 1.49% overhead.

| | | Total CPU time (sec) | Total time (%) |
|---|---|---|---|
| | **Application computation** $comp(Df)$ | 1,407,967.18 | 98.18% |
| **Monitoring** | **Provenance** $prov(Df)$ | 4,259.18 | 0.3% |
| | **Data extraction** $ext(Df)$ | 21,367.60 | 1.49% |
| **Steering** | **Steering point** $s_{point}(Df)$ | 473.24 | 0.03% |
| | **Steering action** $s_{action}(Df)$ | 2.44 | 1.7e-5% |
| | **Total** $c(Df)$ | 1,434,069.64 | 100% |

Any overhead caused by this solution is greatly compensated by the benefits we make available to the user. For example, keeping the registry of the adaptations related to the provenance of the results benefits reproducibility, validation, and interpretation (Challenge 2). Also, observing at runtime that the adaptation reduced the execution time in ten days (Challenge 3) is relevant for further online tunings and result analyses.

## 8. Conclusion

In this paper, we proposed a solution for keeping track of user steering actions in dynamic workflows. We provided a formal definition for steering action and the tracking of parameter tuning in dataflows of workflows. We extended a W3C PROV provenance model for data representation of fine-tuning of parameters, which is very frequent steering action available in several computational steering systems. We also presented DfAdapter, a tool to facilitate scientists to fine-tune parameters online while managing provenance of steering actions for the tunes. DfAdapter works in the same way as visualization libraries like ParaView are used in workflows. Strategic calls to DfAdapter tracking services are inserted at the adaptation service invocations of the user workflow. DfAdapter captures provenance of steering actions and stores in its database, relating with data for workflow execution, dataflow provenance, and especially with strategic domain values, like QoI. The database is available for online data analyses via structured query or graphic interfaces.

We developed a case study of DfAdapter using a real sedimentation simulation dynamic workflow in the Oil and Gas industry, using large and small-scale experiments. By using data captured by DfAdapter, the user could verify which parameters contributed to a reduction of simulation time. Also, the steering data registry enabled the user to verify that tuning specific parameters made it finish successfully. The user could run, for example, Query 2, which integrates data for provenance, performance, and the new fine-tuning data introduced in this paper. The user was able to perform steering actions based on the analysis of the impact of each previous tune as registered by DfAdapter. Without DfAdapter support, fine-tuning could be error prone and compromise the reliability of the results. We also observed that the added overhead for DfAdapter for provenance and steering accounted for less than 1% of total simulation time.

Therefore, in this work we contributed to provenance management and online analysis of user steering actions in the context of putting humans in the loop of dynamic workflows, which is considered a challenge. Thus, in addition to typical uses of provenance data (*i.e.*, result reproducibility, reliability, and validation), we exploit them for online data analysis, supporting users in their decision-making process. In future work, we plan to explore other types of steering actions in workflows.

## ACKNOWLEDGMENTS

This work was partially funded by CAPES, CNPq, FAPERJ and Inria (MUSIC and SciDISC projects), EU H2020 Programme and MCTI/RNP-Brazil (HPC4E grant no. 689772), and performed (for P. Valduriez) in the context of the Computational Biology Institute.